\documentclass[10pt,journal,compsoc]{IEEEtran}

\usepackage{txfonts}
\usepackage[table,xcdraw]{xcolor}

\usepackage{graphicx} 
\usepackage{pdfpages}			
\usepackage{times}              
\usepackage[ruled,vlined]{algorithm2e}
\usepackage{lipsum}
\usepackage{epstopdf}
\usepackage{tabularx}
\usepackage{array}
\usepackage{subfigure}
\usepackage{pbox}

\usepackage{indentfirst} 
\usepackage{setspace}
\usepackage{hyperref}

\usepackage{blindtext}
\usepackage{booktabs}
\usepackage[utf8]{inputenc}
\usepackage{verbatim}
\usepackage{threeparttable}

\usepackage{hyphenat}

\usepackage{amssymb}
\usepackage{xspace}
\usepackage{lineno}
\usepackage{color}

\ifCLASSOPTIONcompsoc
  \usepackage[nocompress]{cite}
\else
  \usepackage{cite}
\fi

\begin{document}

\title{A softwarized perspective of the 5G networks}
\author{
    Kleber Vieira Cardoso,
    Cristiano Bonato Both, 
     Lúcio Rene Prade, \\
     Ciro J. A. Macedo,
     Victor Hugo L. Lopes
     \thanks{Kleber, Ciro, and Victor are with Universidade Federal de Goi\'{a}s (UFG), Brazil. Email: \{kleber, ciro.macedo, victor.lopes\}@inf.ufg.br}
    \thanks{Cristiano and Lúcio are with the University of Vale do Rio dos Sinos (UNISINOS), Brazil. Email: \{cbboth, luciorp\}@unisinos.br}
}

\clearpage\maketitle
\thispagestyle{empty}

\begin{abstract}
The main goal of this article is to present the fundamental theoretical concepts for the tutorial presented in IEEE NetSoft 2020. The article explores the use of software in the 5G system composed of the Radio Access Network (RAN) and the core components, following the standards defined by 3GPP, particularly the Release 15. The article provides a brief overview of mobile cellular networks, including basic concepts, operations, and evolution through the called `generations' of mobile networks. From a software perspective, RAN is presented in the context of 4G and 5G networks, which includes the virtualization and disaggregation concepts. A significant part of the article is dedicated to 5G networks and beyond, focusing on core, \textit{i.e.}, considering the Service-Based Architecture (SBA), due to its relevance and totally softwarized approach. Finally, the article briefly describes the demonstrations presented in IEEE NetSoft 2020, providing the link for the repository that has all material employed in the tutorial.
\end{abstract}
\section{Introduction}\label{sec:introducao}

In the generations leading up to the fifth generation of wireless cellular mobile networks, the supreme effort was made in the massification of access to customers and increasing the transmission rates. However, more than just improving bandwidth and reducing latency, 5G networks~\cite{shafi:17} will allow truly disruptive solutions to emerging in all types of industries~\cite{kaloxylos:18}. Although many countries, such as Brazil, are still discussing and preparing for the deployment of 5G networks, since 2019, several countries (\textit{e.g.}, China, South Korea, and the United States) are already implementing, testing, and effectively using the latest generation of mobile wireless networks. More than 2.4 billion devices are expected to use 5G networks by 2025~\cite{richter:19}. For now, rates close to (and even higher than) 1~Gbps are still the main novelties, but several services, mainly based on Internet of Things (IoT), began to be planned and implemented. Autonomous cars, Industry 4.0, and virtual/augmented/mixed reality are some of the bets for relevant applications and tend to serve as interesting use cases for new network features. However, as in previous generations, it is difficult to predict accurately which applications or services will be adopted. For the time being, the only consensus is that 5G networks, when deployed on a significant scale, will represent a new step for wireless communications and will have a deep impact in the society~\cite{wef:20}.

The fifth-generation represents a remarkable technological leap over the fourth generation, introducing significant hardware and, above all, considerable software innovations. It is essential to clarify that this leap hides the various intermediate steps that comprise it. In the telecommunications industry, including for advertising reasons, a significant amount of technological developments and innovations accumulate before defining a generation, which was no different at 5G networks. As an example, the Releases provided by the 3GPP (3rd Generation Partnership Project) that is one of the leading organizations for standardizing wireless mobile networks. The 4G/LTE networks (Long-Term Evolution) were introduced in 2008 by Release 8~\cite{3GPP-rel08:19} and, since then, have received several updates (represented by new Releases) until the 5G networks were officially introduced by Release 15~\cite {3gpp:rel15nr21.915} in 2018. On the other hand, there are technical factors that also support the definition of each generation. In the 5G system, the radio spectrum usage is expanding to hundreds of MHz or some units of GHz, but it also going to frequency bands of tens of GHz. Moreover, 5G networks consolidate an intense process of softwarization that stands out for the adoption of cloud systems and technologies such as virtualization~\cite{abdelwahab:16}, software-defined networks~\cite{chen:15}, network slicing~\cite{foukas2017network}, and Service-Based Architecture (SBA)~\cite{mademann:18}.

\begin{figure}[htb]
\centering
    \subfigure[5G scenarios]
    {\label{subfig:cenarios_5g}
    \includegraphics[width=.48\textwidth]{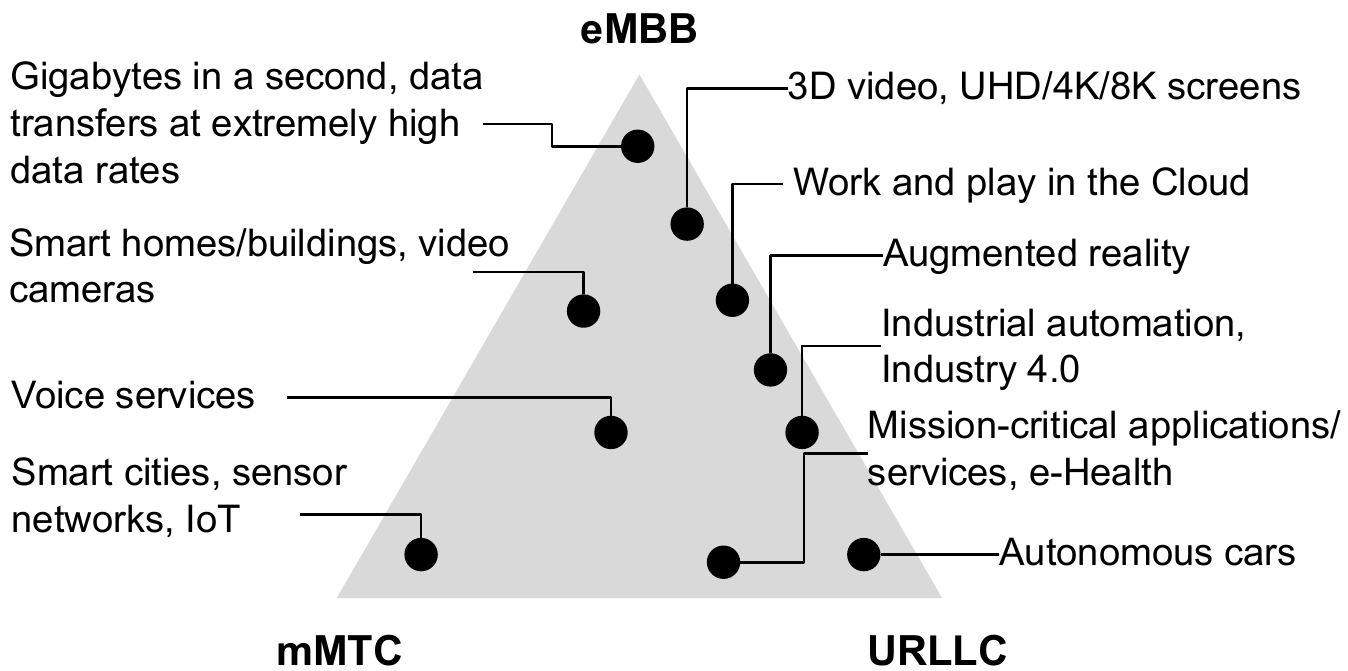}}
    \hfil
    \subfigure[Capabilities of the scenarios]
    {\label{subfig:capacidades_cenarios_5g}
    \includegraphics[width=.48\textwidth]{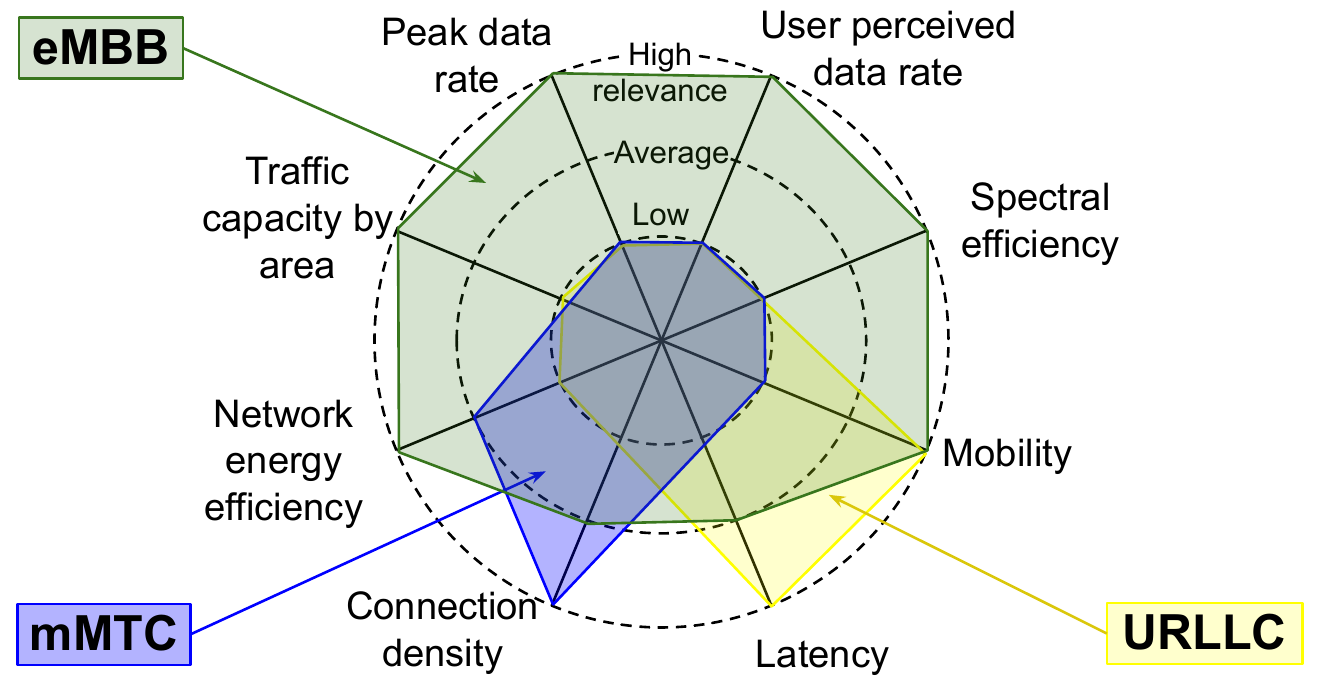}}
    \caption{Scenarios (or requirements) defined for 5G networks (left) and capabilities offered in these scenarios (right).}
\label{fig:redes_5g}
\end{figure}

In addition to the technological aspects, the 5G networks introduce changes in the companies' business model. Traditionally, operators of wireless mobile communication systems have focused on end-users as the main sources of revenue. In 5G networks, the intention is to expand (or even migrate) the focus to have industries as primary customers~\cite{palattella:16,lema:17}. Three main requirements (or scenarios) have been defined to support this expansion of the telecommunications companies' business model: enhanced mobile broadband (eMBB), ultra-reliable low-latency communications (URLLC), and massive machine-type communications (mMTC). These scenarios are illustrated in Fig.~\ref{subfig:cenarios_5g}, which was introduced by ITU (International Telecommunication Union)~\cite{itu:15}, in 2015, as part of the vision of what 5G networks would be. This illustration, which presents the leading capabilities of each scenario, is widely used to try to summarize some relevant properties of 5G networks.

In order to meet the requirements of the different scenarios, the concepts of network software began to be adopted, both by academia and industry, to make the development of 5G networks (or systems) viable in recent years and also their evolution~\cite{foukas2017network}. The main reason for the intensive introduction of software in 5G systems is to extend the flexibility to the mobile network architecture, supporting different requirements in challenging scenarios emerging in the digitalized society. Moreover, with the network softwarization, it is possible to enjoy of traditional cloud computing, as well as its smaller-scale variants, \textit{i.e.}, fog and edge. In this context, radio access and core network functions are now implemented using the advantages of cloud computing, such as abundant storage, large-scale processing, elasticity, etc. In addition to enable new services, in the long run, the softwarization is expected to reduce the costs of deploying and operating 5G systems.

The remainder of this article is structured as follows. Section~\ref{sec:redesMoveis} introduces basic concepts related to mobile network cellular networks and the operation of this type of system. This section presents a brief historic review of the generations until the forth. Section~\ref{sec:5G} is focused in the fifth generation and its evolution. Some key technologies, mainly related to software, are presented in this section. In the Section~\ref{sec:RAN}, the Radio Access Network (RAN) is presented in more detail and some aspects of softwarization process is discussed. Section~\ref{sec:core} describe the 5G core and most of the several functions that compose the new architecture. In the Section~\ref{sec:integracao}, some aspects of the integration of a 5G system and non-3GPP access networks are presented. This section also discuss in more detail the integration of a 5G core and LoRa access networks. Finally, Section~\ref{sec:conclusao} summarizes the contributions expected in the Releases 16 and 17, mainly in the software context.
\section{Mobile cellular networks until the forth generation}\label{sec:redesMoveis}

While the word `cellular' has become almost redundant nowadays when referring to mobile networks, it summarizes several concepts and innovations. A mobile radio (telephone) service was initially provided by high-power transmitters/receivers that could communicate directly to each other until a radius of about 80~km. These devices were heavy, expensive, consumed a lot of energy, and, as expected, had no support for data communication. The introduction of the cellular radio changed the scenario drastically, allowing the development of small, cheap, energy-efficient, and, later, data-oriented communication devices. Since the first commercial mobile cellular network, in 1979, until nowadays, several new technologies have been introduced in the communication systems, but also several concepts remain the same or are very similar. In this section, we present relevant concepts of mobile networks in general and the ones related to cellular networks. Additionally, we describe the history of the mobile cellular networks very briefly through the `generations', starting from the first until the fourth, with most of the attention to the fourth generation. The fifth-generation and beyond have the next section dedicated to them.

\subsection{Basic concepts}

\begin{figure}[htb]
\centering
    \subfigure[Ideal]
    {\label{subfig:cell_net_ideal}
    \includegraphics[width=.32\textwidth]{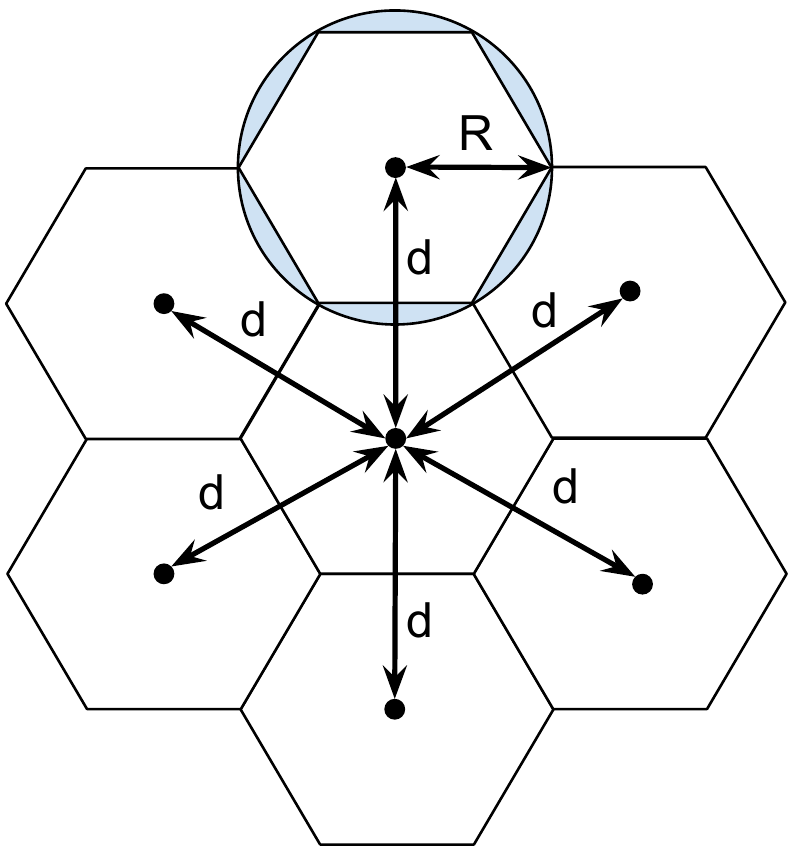}}
    \hfil
    \subfigure[Practical (positions randomly generated in GNU Octave)]
    {\label{subfig:cell_net_practice}
    \includegraphics[width=.32\textwidth]{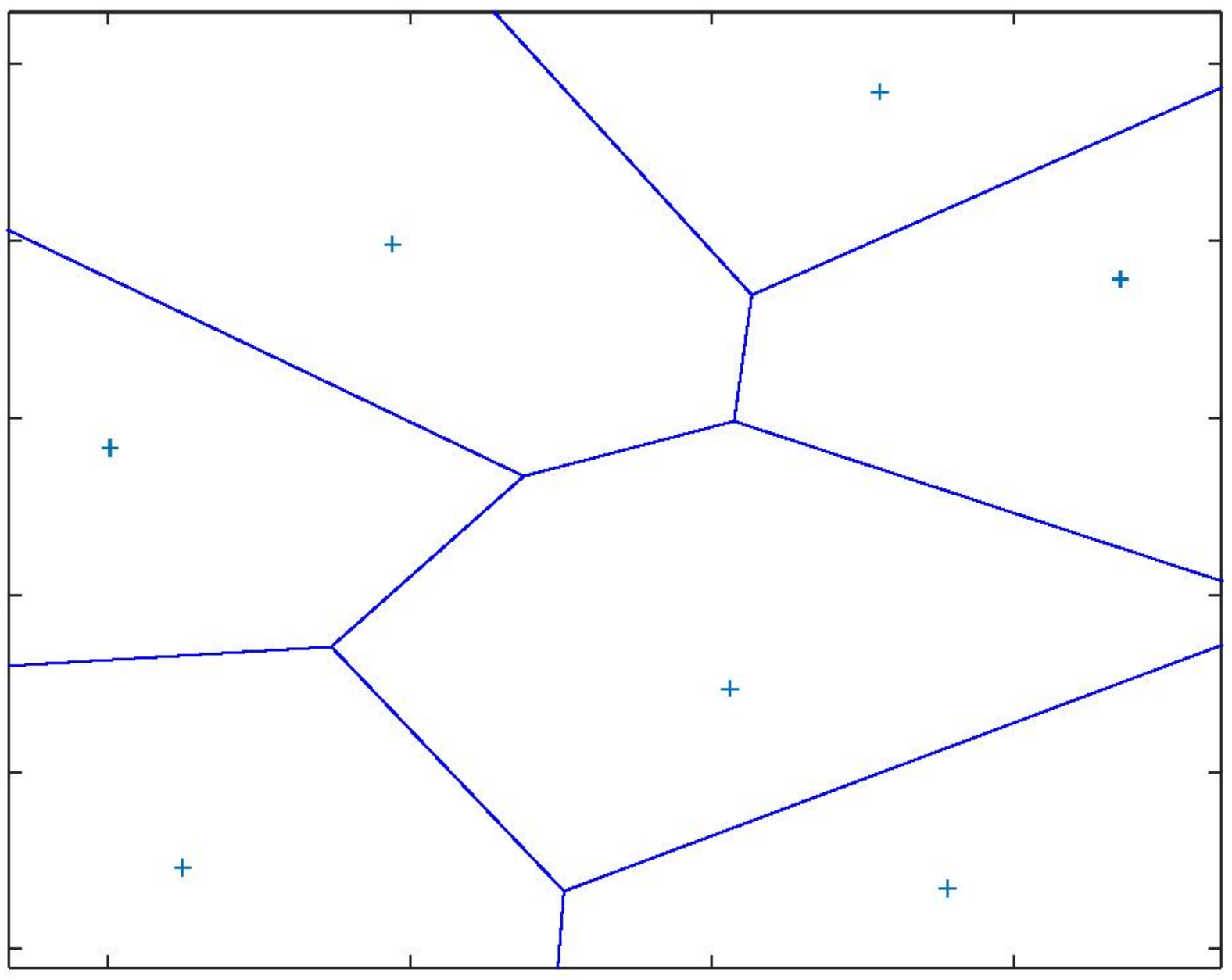}}
    \caption{Representation of a cellular network.}
\label{fig:cell_net}
\end{figure}

The fundamental idea of a cellular network is the use of multiple low-power transceivers (transmitters+receivers) to offer communication to mobile devices. Actually, the power of these transceivers is low compared with their predecessors, but it is very high compared with mobile devices. Traditionally, these transceivers are mounted in a mast or tower, at an elevated point, and have a control unit. This configuration is named as a Base Station or BS, which is responsible for covering an area using a band of frequencies. Assuming the most traditional positioning in the center of the coverage area, the area's conceptual representation as a hexagon was the appropriate choice because the distance between the centers of all adjacent areas is the same (equal to $\sqrt{3}R$). Therefore, a mobile cellular network can be represented as a collection of the neighboring regions, each resembling a cell, as illustrated in Fig.~\ref{subfig:cell_net_ideal}. In practice, the environmental characteristics and the propagation conditions tend to degenerate the mobile cellular network into something more similar to a Voronoi diagram or tessellation, as illustrated in Fig.~\ref{subfig:cell_net_practice}.

The deployment of the BSs is generally defined by a previous study that takes into consideration factors as topographical characteristics, signal propagation conditions, limitations on siting antennas, and demand of the users. While in-site measurements are common, most of the study is generally performed with propagation models that can estimate the signal power received along the whole covered are. These models try to represent in a very compact way the complex propagation dynamics of the wireless signal. Table~\ref{tab:prop_models} lists a small set of models available in the literature and their frequency ranges.

\begin{table}[htb]
\centering
\begin{tabular}{ll}
\hline
\textbf{Model} & \textbf{Frequency range}  \\
\hline
Okumura-Hata & 150--1500~MHz\\
COST 231 -- Walfish-Ikegami & 800--2000~MHz\\ 
ITU-R P.529 & 700--3500~MHz \\ \hline
\end{tabular} 
\caption{Examples of propagation models.}
\label{tab:prop_models}
\end{table}

Traditionally, a mobile cellular network operates in licensed bands which must be acquired according to the specific set of rules of each country and may be very expensive. Therefore, a lot of the effort is devoted to managing the frequencies available for communication properly. This management involves several coordinated approaches to seek the efficient usage of the Radio Frequency (RF) spectrum, which include antenna schemes, power control, interference control, frequency reuse, among others.

While all these concepts are still relevant in modern cellular networks, new concepts and technologies such as Multiple-Input Multiple-Output (MIMO), beamforming, and millimeter-wave communications have a significant impact. For example, beamforming makes it more complex to define cell coverage and also to identify interference. Communications in the millimeter-wave bands (\textit{i.e.}, above 24~GHz) have very different propagation characteristics than the traditional sub-6~GHz employed along decades in the mobile-cellular networks.

\subsection{Basic operation of a cellular system}

\begin{figure}[htb]
 \begin{center}
\includegraphics[width=0.5\textwidth]{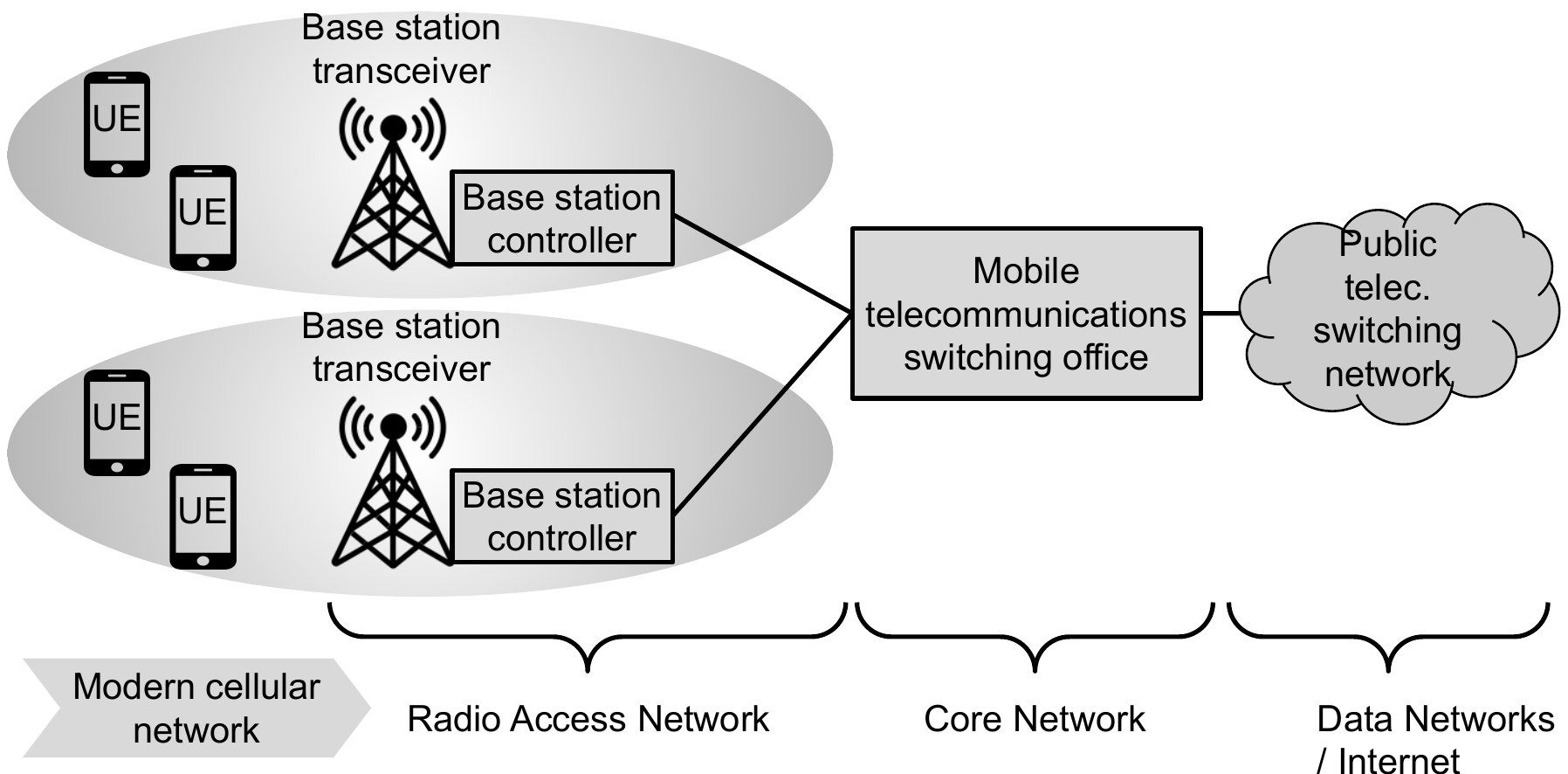}
  \end{center}
\caption{Overview of a basic cellular system.}
\label{fig:cell_system}
\end{figure}

In its original configuration, a cellular system could be summarized into the following elements: Mobile Telecommunications Switching Office (MTSO), BS, and mobile unit or User Equipment (UE). Fig.~\ref{fig:cell_system} illustrates a basic cellular system. In a modern cellular system, the MTSO can be identified as the core (network), while the collection of BSs is named as Radio Access Network (RAN). Actually, several other elements are composing the core and RAN, as we describe later in this and the following sections.

The BS controller handles the communication process between the UE and the rest of the network. Each BS may simultaneously serve multiple UEs under its coverage area. One MTSO serves multiple BSs, being connected generally through wired links, but wireless links are also common. Traditionally, all communications between UEs are established through a BS, even when they are close to each other. The concepts, technologies, and standards for direct communication (known as device-to-device or D2D~\cite{kar:18}) between UEs have already been introduced in cellular networks. However, when its effective adoption begins, it remains uncertain. An MTSO connects to the public telephone or telecommunications network, offering interconnection between its UEs and the outside networks. Several other management tasks are performed by an MTSO to assist its UEs, such as channel (de)allocation, handoff, paging, monitoring, and billing.

The channel (de)allocation may occur in different situations, for example, when UE is turned on or off, when UE moves from one BS coverage to another, if UE has a signal too weak, among others. The handoff process consists of the coordinated transfer of UE communication while moving between neighboring BSs, involving signal monitoring and channel allocation and deallocation. The paging process consists of finding a UE that has not been communicating for some time and so its location is unknown. Generally, the paging process is started by an incoming communication to UE. In addition to the wireless signal quality, the cellular network also monitors the traffic of UEs to charge the users properly. All these procedures remain in modern mobile networks, naturally, more focused on data instead of voice. Furthermore, technological advances imply updating some concepts. For example, instead of channel allocation and deallocation, modern mobile networks may allocate and deallocate Physical Resource Blocks (PRBs) and millimeter-wave beams, and even edge computing resources.

Similar to the traditional telecommunication systems, in a cellular system, the only concerns of a user are to initiate or answer a communication. The network system transparently manages everything, \textit{i.e.}, the user does not need to worry about configuration, reconfiguration, tuning, or any related task. Similar to traditional telecommunication systems, a cellular system has two types of channels between UEs and BSs: control and traffic channels. Control channels are used to exchange information related to control and management tasks (\textit{e.g.}, UE initialization, handoff, paging). Traffic channels are used to transport (voice or data) communications. In modern cellular networks, a more generic nomenclature is adopted, control plane, and user plane, but the meaning is very similar.

\subsection{Evolution of the mobile cellular networks}

By employing the arguable separation in `generations', the characteristics of the mobile cellular networks, until the fourth generation, can be loosely summarized by the information presented in Table~\ref{tab:geracoes}. A detailed description of the history of mobile cellular networks can be found in~\cite{stallings2014data,schiller2003mobile}, since our concerns are only the main characteristics of each generation and their relationship with the softwarization process. Naturally, the fourth generation deserves special attention since it is the first one designed to have data and software as relevant elements and also due to its present market share.

\begin{table*}[tbh]
\centering
\begin{threeparttable}
\begin{tabular}{lcccc}
\hline
\textbf{}      & \textbf{1G}   & \textbf{2G} & \textbf{3G}                         & \textbf{4G}    \\ \hline
Deployment  & $\approx1980$          & $\approx1991$        & $\approx1999$     & $\approx2009$  \\ 
Main services       & Analog voice & Digital voice, SMS & Digital voice, data packets & IP packets \\ 
Data rate  & 1.9 kbps      & 14.4 kbps   & 384 kbps                            & 200 Mbps       \\ 
RAN &   \begin{tabular}[c]{@{}c@{}}AMPS, TACS, NMT,\\ C-450, TMA, RTM\end{tabular}   &   \begin{tabular}[c]{@{}c@{}}GSM, GPRS, EDGE,\\PDC, IS-95, IS-136\end{tabular} & \begin{tabular}[c]{@{}c@{}}UMTS (W-CDMA),\\ CDMA2000, TD-SCDMA \end{tabular}  &  LTE, WiMAX
\\ 
Core & PSTN (SS7)        & PSTN (SS7, ISDN)   & PSTN, ATM, IP     & IP network (EPC)            \\ 
3GPP initial standard           &      -         &       -      &      Release 99 &  Release 8         \\
Global market share (by 2019)\tnote{1} & 0\% & 23\% & 25\% & 52\% \\ \hline
\end{tabular}
\begin{tablenotes}
\item[1] \url{https://www.statista.com/statistics/740442/worldwide-share-of-mobile-telecommunication-technology/}
\end{tablenotes}
\caption{Characteristics of the generations of the cellular networks.}
\label{tab:geracoes}
\end{threeparttable}
\end{table*}

Nowadays, 1G is the only one without commercial deployments in operation, while the other generations still exhibit a significant presence. For example, 2G, with the lowest participation, still has more than 20\% of the market share. The substantial investment in the legacy hardware and new opportunities, such as the support for IoT communications, can partially explain the long-living of 2G and 3G networks. Until the 3G, there was a considerable fragmentation in the radio access technologies, which is illustrated by the several standards available. A large number of players and the trade-offs of the competing technologies are some characteristics of the maturity level of the cellular networks at that time and helps to explain the fragmentation scenario. By the end of 2G and the beginning of 3G, the 3GPP initiative was introduced, having great success in 3G and 4G. The fourth-generation is the first focused on data communication. It offers, by design, proper support for the IP (Internet Protocol) stack and notably higher data rate in comparison with the previous generation.

\subsubsection*{First generation -- 1G}

The first commercially deployed cellular networks, now labeled as 1G, were an extension of the Public Switched Telephone Network (PSTN). Therefore, the focus was on analog voice communication, without any concern about data communication. Later, the network infrastructure was adapted to transport data but achieving only very low rates. The software in this generation was proprietary, low-level, and embedded in specific hardware. The network control was digital, but voice communication was analog. Despite the several limitations, the first-generation cellular networks were a great success, motivating the evolution and giving rise to a high-valuable and strategic industry.

\subsubsection*{Second generation -- 2G}

Compared with 1G, the second-generation cellular networks provided higher-quality signals, higher data rates, and greater capacity. 2G introduced digital traffic channels, \textit{i.e.}, 2G systems support digital data natively, and voice is digitally encoded before transmitting. Since all information becomes digital, 2G systems started offering support for encrypting both user and control traffic. The digital traffic also allows the adoption of error detection and correction techniques, which improves the quality of voice communication. 2G systems also introduced new services, such as the Short Message Service (SMS), that was very successful. Time-Division Multiple Access (TDMA) and Code Division Multiple Access (CDMA) were also introduced, improving efficiency in the use of the RF spectrum. While bit rates offered by 2G networks are too slow for many modern applications, they are suited for a wide range of IoT demands. This suited possibility has motivated infrastructure operators around the world to extend the life of 2G systems and vendors to revisit some products to improve relevant performance metrics for IoT, such as energy efficiency.

The software in 2G systems already had an important role, being responsible for the critical control and management tasks, including monitoring and billing. However, proprietary technologies and specialized appliances, \textit{i.e.}, combined hardware and software, were dominant. On the other hand, the adoption of solutions based on standards for open systems, such as the Telecommunications Management Network (TMN) defined by ITU-T, may be seen as initial steps to a more advanced softwarization process in the telecommunication area. In 2G systems, Global System for Mobile Communications (GSM) had great success, and, nowadays, some open-source software projects implement enough GSM features to have a basic operational cellular network. For example, OpenBTS and YateBTS\footnote{OpenBTS: \url{http://openbts.org/}; YateBTS: \url{https://yatebts.com/}} offer open-source software for the Base Transceiver Station (BTS), while OsmocomBB\footnote{https://osmocom.org/projects/baseband} implemented an open-source GSM baseband software.

\subsubsection*{Third generation -- 3G}

The benefits introduced by the second generation expanded the global deployment and the interest of the general public in mobile networks. In the early 90s, the ITU requested proposals for radio transmission technologies as part of the International Mobile Telecommunications (IMT) 2000 program. This program planned to have a universal global system, but after many discussions and disputes about patents, a family of 3G standards was adopted. 3GPP, for the Universal Mobile Telecommunications System (UMTS), and 3GPP2\footnote{3GPP2 is not an evolution of 3GPP, but a different standardization body.}, for the CDMA2000, became the main driving forces in the third generation's standardization process. Release 99 is the initial set of 3G UMTS standards published by 3GPP and the last one numbered according to the standardization year (or close since it was finished in 2000). The next is Release 4 that describes the UMTS all-IP Core Network.

In comparison with 2G, the third-generation cellular networks increased the data rates, improved the quality of voice communication, improved the efficiency in the use of the RF spectrum, enhanced the integration with IP networks, and enhanced the support for new services. However, 3G still maintains backward compatibility with several previous technologies, \textit{e.g.}, circuit switch is always supported. Although incompatible variants multiple have been adopted, CDMA was largely evolved in 3G and became the dominant technology for wireless communication in that generation. The software in 3G systems become even more relevant than in the previous generation since the infrastructure is more complex, and data communication becomes more relevant. However, proprietary software and vendor-specific appliances protected by endless patents increase the costs and delay innovation. On the other hand, in the first years of the millennium, mobile cellular networks were already essential, so the evolution of these systems became unstoppable.

Similar to 2G, 3G also may have its life extended due to IoT and also due to the more recent and significant investments. During the deployment of 3G, the telecommunication sector faced an economic world crisis and expensive auctions for obtaining the RF spectrum. UMTS was the most successful standard in 3G and, similar to GSM, has some open-source software, provided by the same developer groups. OpenBTS-UMTS\footnote{\url{https://github.com/RangeNetworks/OpenBTS-UMTS}} implements the basic functionalities according to the Release 99, using OpenBTS as a framework. The project Cellular Network Infrastructure\footnote{\url{https://projects.osmocom.org/projects/cellular-infrastructure/}} from Osmocom offers a complete implementation that allows for deploying an operational 3G system.

\subsubsection*{Fourth generation -- 4G}

\begin{figure}[htb]
 \begin{center}
    \includegraphics[width=0.5\textwidth]{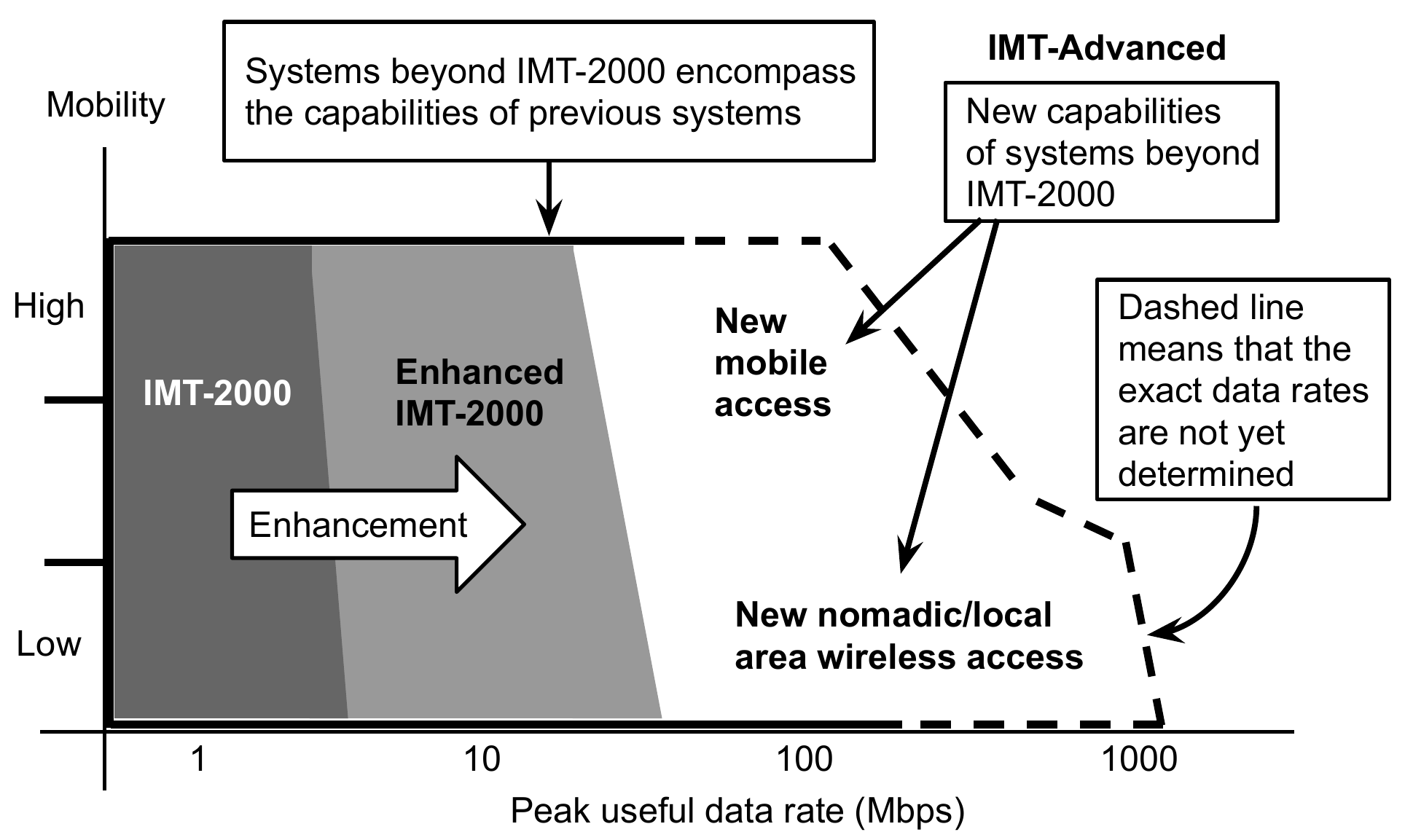}
  \end{center}
\caption{IMT-2000 capabilities and systems beyond IMT-2000.}
\label{fig:imt-adv}
\end{figure}

The first billion of mobile subscribers was passed in 2002, the second billion in 2005, the third billion in 2007, the fourth billion by the end of 2008, and the fifth billion in 2010~\cite{holma2011lte}. During that decade, the volume of data traffic increased notably, until the global mobile data traffic surpassed voice by the end of 2009\footnote{\url{https://www.ericsson.com/en/press-releases/2010/3/mobile-data-traffic-surpasses-voice}}. Therefore, the plans for evolving the mobile cellular networks into the fourth generation needed to be aggressive to catch up with the demand and data-oriented characteristic. In 2002, the work on IMT-Advanced started by seeking to define the vision and requirements for the next generation of mobile cellular networks. The ``VAN diagram'', introduced by the ITU~\cite{itu:03} and replicated in Fig.~\ref{fig:imt-adv}, summarizes some initial ideas for the fourth generation. In addition to the very high data rates for both mobile and fixed/nomadic users, the IMT-Advanced also identified several other requirements, including the following:
\begin{itemize}
    \item Employ an all-IP packet switched network.
    \item Dynamically share the network resources to support more simultaneous users per cell.
    \item Support smooth handovers across heterogeneous networks.
    \item Offer high quality of service for multimedia applications.
\end{itemize}

In 2009, only two candidate proposals for IMT-Advanced were submitted to ITU: 3GPP LTE-Advanced (LTE-A) and IEEE 802.16m (Mobile WiMAX Release 2 or WirelessMAN-Advanced). In the following years, several deployments of LTE and WiMAX started around the world. Initially, these deployments were not IMT-Advanced, \textit{i.e.}, they were not precisely 4G networks, but they could be easily upgraded from LTE to LTE-A and from WiMAX 1.5 to WiMAX 2.0. LTE-A and WiMAX 2.0 are similar in terms of both performance and some technologies. For example, both are based on the use of Orthogonal Frequency-Division Multiple Access (OFDMA) to support multiple access to network resources. However, the OFDMA approaches are not compatible, and there are other relevant differences, such as the backward compatibility and core network. Along this last decade, the differences favored the 3GPP LTE-A as the leading 4G solution. On the other hand, WiMAX had an important role of being a high-quality and aggressive competitor and so motivated the 3GPP consortium to notably improve the initial LTE. Nowadays, WiMAX is being employed in important niches, such as data communications and information sharing on the airport surface, wireless broadband communications for Smart Grid network applications, specialized private networks for Oil and Gas companies, and wireless backhaul in general.

Table~\ref{tab:4g_rel} summarizes the main content introduced in each 3GPP release related to the fourth generation\footnote{\url{https://www.cablefree.net/wirelesstechnology/4glte/overview-of-lte-3gpp-releases/}}. In Release 8, the Evolved Packet System (EPS) is primarily defined, being composed of  Evolved Universal Terrestrial Access Network (E-UTRAN) and System Architecture Evolution (SAE). Relevant technologies, such as OFDMA and Multiple Input Multiple Output (MIMO), are also added to the standard in Release 8. E-UTRAN is known as Long Term Evolution (LTE) while SAE is generally referenced as Evolved Packet Core (EPC). Release 9 introduced the complete integration of the Home eNodeB (HeNB), evolved the Self-Organizing Networks (SON) and the multimedia broadcast and multicast service (eMBMS), and also added new spectrum bands (\textit{e.g.}, 800 MHz and 1500 MHz) for LTE operation, and interoperability between LTE, WiMAX, and UMTS. LTE-Advanced, introduced in Release 10, can be considered as a toolbox of features that can be flexibly implemented on top of LTE, including carrier aggregation (up to 100 MHz), MIMO evolution (up to 8x8 in downlink and 4x4 in uplink), relay nodes, and enhanced Inter-Cell Interference Coordination (eICIC).

\begin{table}[htb]
\centering
\begin{tabular}{lcl}
\hline
\textbf{Year}      & \textbf{Release}   & \textbf{Content (extremely summarized)}  \\ \hline
2008 & 8 & LTE is introduced \\
2009 & 9 & Enhancements to LTE \\
2011 & 10 & LTE-Advanced \\
2012 & 11 & Enhancements to LTE-Advanced \\
2015 & 12 & Further enhancements to LTE-Advanced \\
2016 & 13 & Matching the increasing throughput demand \\
2017 & 14 & First steps into 5G standardization \\ \hline
\end{tabular} 
\caption{Main content introduced in the 4G-related 3GPP releases.}
\label{tab:4g_rel}
\end{table}

After accomplishing the requirements of IMT-Advanced in Release 10, 3GPP continued the standardization of some enhancements from the previous features and also of new functionalities and services. For example, Release 11 introduced carrier aggregation enhancements, further enhanced ICIC (FeICIC), further SON enhancements, RAN Enhancements for diverse data applications, Coordinated Multipoint (CoMP), advanced IP interconnection of services, among others. A priority in Release 12 was the use of LTE technology for emergency and security services. Other notable features added in this release included small cells and network densification, device-to-device (D2D) communications, Machine Type Communications (MTC), and WiFi integration into mobile operator's offerings. In Release 13, 3GPP continued to carrier aggregation to large aggregate numbers of carriers in different bands and identified Beamforming and MIMO as key technologies to address the future capacity demand. Release 13 also included: LTE in unlicensed spectrum (also known as Licensed-Assisted Access), enhancements for MTC, enhancements for D2D, and indoor positioning.

The Release 14 introduced several enhancements and new technologies, including improvements in the Mission Critical aspects (introduction of video and data services), the introduction of Vehicle-to-Everything (V2X) aspects, advancements in the Cellular Internet of Things (CIoT) aspects, improvements in the radio interface (in particular by enhancing the coordination with WLAN and unlicensed spectrum), enhancements for TV service, multimedia priority service modifications, eMBMS enhancements, among others. On the other hand, the discussion about 5G had a lot of attention in this release, and of the novelties in Release 14 were introduced to contribute to the transition to the new generation. For example, Control and User Plane Separation (CUPS) and enhanced DECOR (eDECOR) are considered key features that help to meet the way towards 5G\footnote{\url{https://www.3gpp.org/news-events/1822-sa-rel-14}}. We present additional information about these two features in the next section, \textit{i.e.}, in the context of the fifth-generation. 

\begin{figure}[htb]
 \begin{center}
\includegraphics[width=0.5\textwidth]{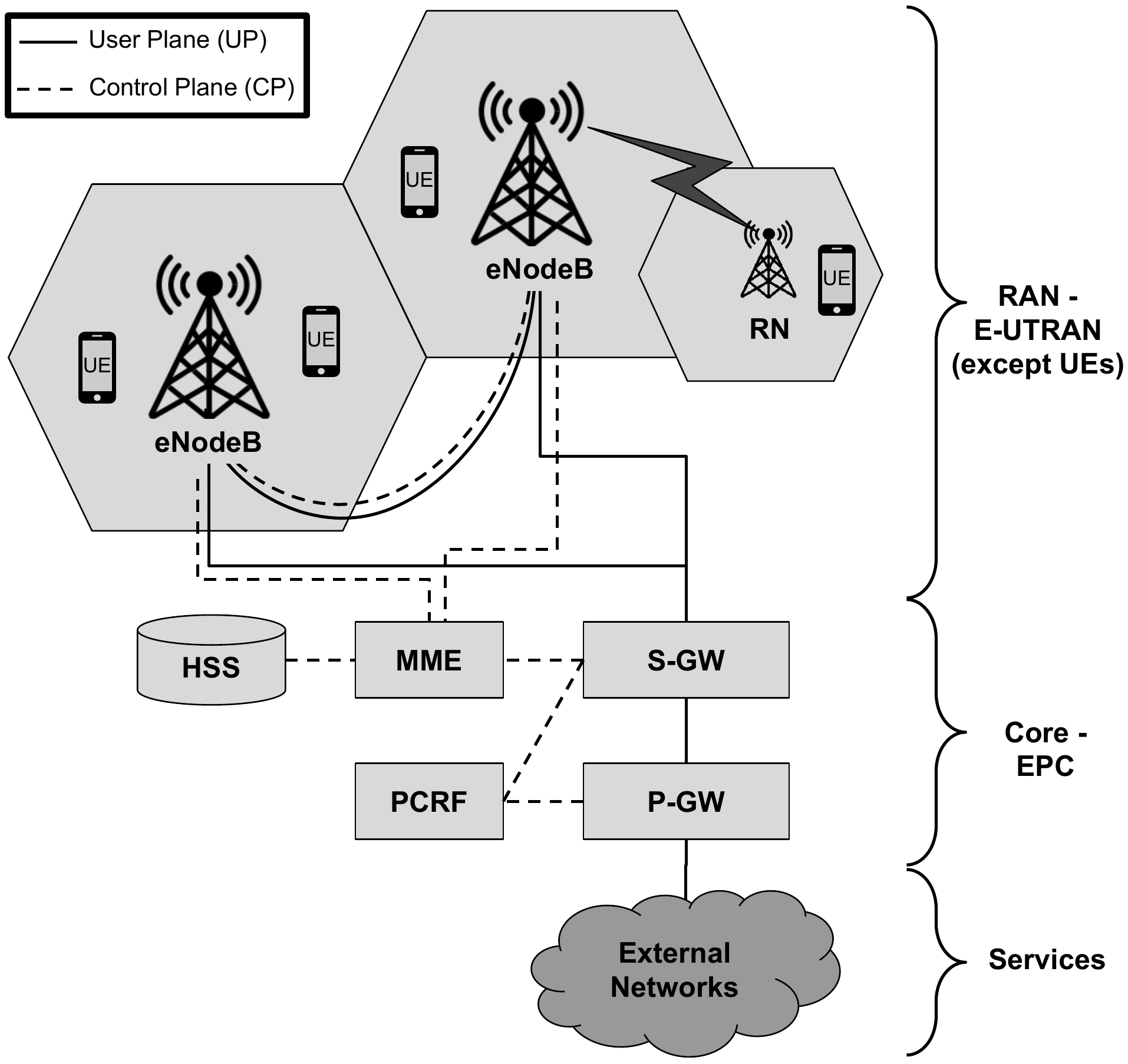}
  \end{center}
\caption{System architecture of EPS with E-UTRAN only.}
\label{fig:lte-adv}
\end{figure}

In terms of the software perspective, 4G can be considered a noticeable evolution compared with the previous generations. Several incentives (both positive and negative) have been driving the softwarization process, including open and easily accessible standards, the contribution of multiple communities in the standardization process, pressure for lowing costs, pressure for facilitating innovation, pressure for multi-vendor integration, among others. On the other hand, some of the most powerful software paradigms, such as virtualization, software-defined networking, microservices and, other cloud computing-related ones, are still incipient or absent in the 4G systems. While several basic functionalities of RAN and core can be implemented in general propose hardware, there are still many advanced features that are only available with the help of specialized hardware (\textit{e.g.}, Field-Programmable Gate Array -- FPGA) and protected by patents and copyrights (e\textit{.g.}, Intellectual Property -- IP). Actually, patents and copyrights are not the issues by themselves, but the vendor lock-in and barriers to innovation generally derived from them. Over the last decade, several open-source projects related to 4G have been initiated. Many of these projects were discontinued, mainly those related to WiMAX, but there are projects with active communities and supporters. For example, OpenAirInterface\footnote{\url{https://www.openairinterface.org/}} and srsLTE\footnote{\url{https://github.com/srsLTE/srsLTE}} offer open-source code for fully-functional UE, eNodeB (the central element of a 4G RAN), and EPC (core) to the 3GPP Release 10 or posterior. Before finishing this section, in the following, we briefly describe the LTE-Advanced, \textit{i.e.}, the leading representative of a 4G system.\\
\\
\textbf{LTE-Advanced}\\
\\
Fig.~\ref{fig:lte-adv} illustrates the EPS with only E-UTRAN as the access network, \textit{i.e.}, 2G and 3G access networks, are not represented in this figure. E-UTRAN is concentrated on the evolved Node B (eNodeB) in which all radio functionality is collapsed. As a network, E-UTRAN is simply a mesh of eNodeBs connected to neighboring eNodeBs. However, as we describe in Subsection~\ref{subsec:CRAN}, in a softwarized RAN, the eNodeB can be decomposed in two parts (Baseband Unit -- BBU -- and Remote Radio Head -- RRH), bringing potential benefits such as resource pooling and energy efficiency. 

Traditionally, any cellular network may experience reduced data rates near the edge of its cells due to lower signal levels and higher interference levels. In this context, an optional (but useful) element of the RAN is the Relay Node (RN) that has a reduced radius of operation compared with an eNodeB. Additionally, an RN is more straightforward than an eNodeB, and so it can be more efficient in this context, working as an intermediary between the eNodeB and UEs. On the other hand, an RN is not simply a signal repeater, since it receives, demodulates, decodes the data, applies error correction, and transmits a new signal.

A fundamental change in the architecture of the 4G core networks is the absence of circuit switching, \textit{i.e.}, the EPC does
not have direct connectivity to traditional circuit-switched networks, such as Integrated Services Digital Network (ISDN) or PSTN. As a consequence, voice communication is transported over IP or, more recently, over LTE until the core. The main components of the EPC are the following:

\begin{itemize}
\item Mobility Management Entity (MME) -- This is the main control element in the EPC and operates only in the Control Plane (CP). The MME is involved in authentication, security, mobility management (tracking, paging, handover), management of subscription profile, and service connectivity (bearer setup). Since the first UE register to the network, MME is involved in the authentication process that can be repeated later, including periodically. MME is also responsible for security measures such as ciphering, generation of integrity protection keys, and allocation of temporary identity to the UE. Moreover, to store and update UE-related data, mobility management involves the MME participating in the control signaling with eNodeBs and S-GWs, eventually, with other MMEs. During the time that the MME is serving a UE, it stores a temporary copy of the subscriber profile associated with UE. This profile contains the information that MME needs to adequately assist UE, for example, to set up the bearers to transport the services requested by UE.

\item Home Subscriber Server (HSS) -- This is the data repository for all permanent user data, such as a master copy of the subscriber profile and the permanent key used to calculate the authentication vectors. Some temporary data may be stored in HSS, such as the location of the user in the level of the visited network (\textit{i.e.}, when UE changes of MME) and identities of those P-GWs in use if the support for mobility between non-3GPP access networks is active.

\item Policy and Charging Rules Function (PCRF) -- This network element is responsible for Policy and Charging Control (PCC). Basically, PCRF makes decisions on how to handle the services in terms of QoS and provides the necessary information to the P-GW and, if applicable, also to the S-GW, so that appropriate bearers and policing can be set up.

\item Packet Data Network (PDN) Gateway (P-GW) -- This is the interconnection point between EPC and external networks (\textit{e.g.}, the Internet). P-GW performs traffic shaping and filtering functions according to policies defined for each UE and service. P-GW also collects and reports the related charging information and, typically, allocates the IP address to UEs.

\item Serving Gateway (S-GW) -- This is the interconnection point between RAN and EPC. The main functions of S-GW are User Plane (UP) tunnel management and switching. S-GW is minimally involved in control functions, being responsible for its resources that are allocated based on requests from MME, P-GW, or PCRF.
\end{itemize}
\section{5G networks and beyond}
\label{sec:5G}

The evolution of mobile cellular networks has always sought to reach a ubiquitous high capacity radio, \textit{i.e.}, with increasingly lower levels of latency and higher transmission capacity. At first, the arrival of 5G adds to this evolution the opening of a path for IoT and intelligent communication technologies, exploring new forms of connectivity between Vehicles-to-Infrastructure (V2I), vehicle-to-vehicle (V2V), and Device-to-Device (D2D). However, this significant and widely publicized evolution hides the real 5G revolution, which we describe below.

5G is being developed for the innovation and evolution of cellular systems. This generation is the most dynamic and flexible mobile networks ever designed, making extensive use of cloud-native applications and network core. Moreover, 5G can shaped during its development to absorb new evolutionary leaps within the same pattern. This characteristic aims to accompany the exponential evolution of the leading technologies, such as artificial intelligence and automation. To achieve this maturity, 5G was designed to make extensive use of the virtualization of services and microservices. At first, 5G will be operated by a hybrid network, in which convergence, integration, and coexistence with legacy systems are inevitable. However, in the future, the 5G system will be predominantly composed of virtual network functions.

\begin{figure}[htb]  
  \begin{center}
    \includegraphics[width=0.45\textwidth]{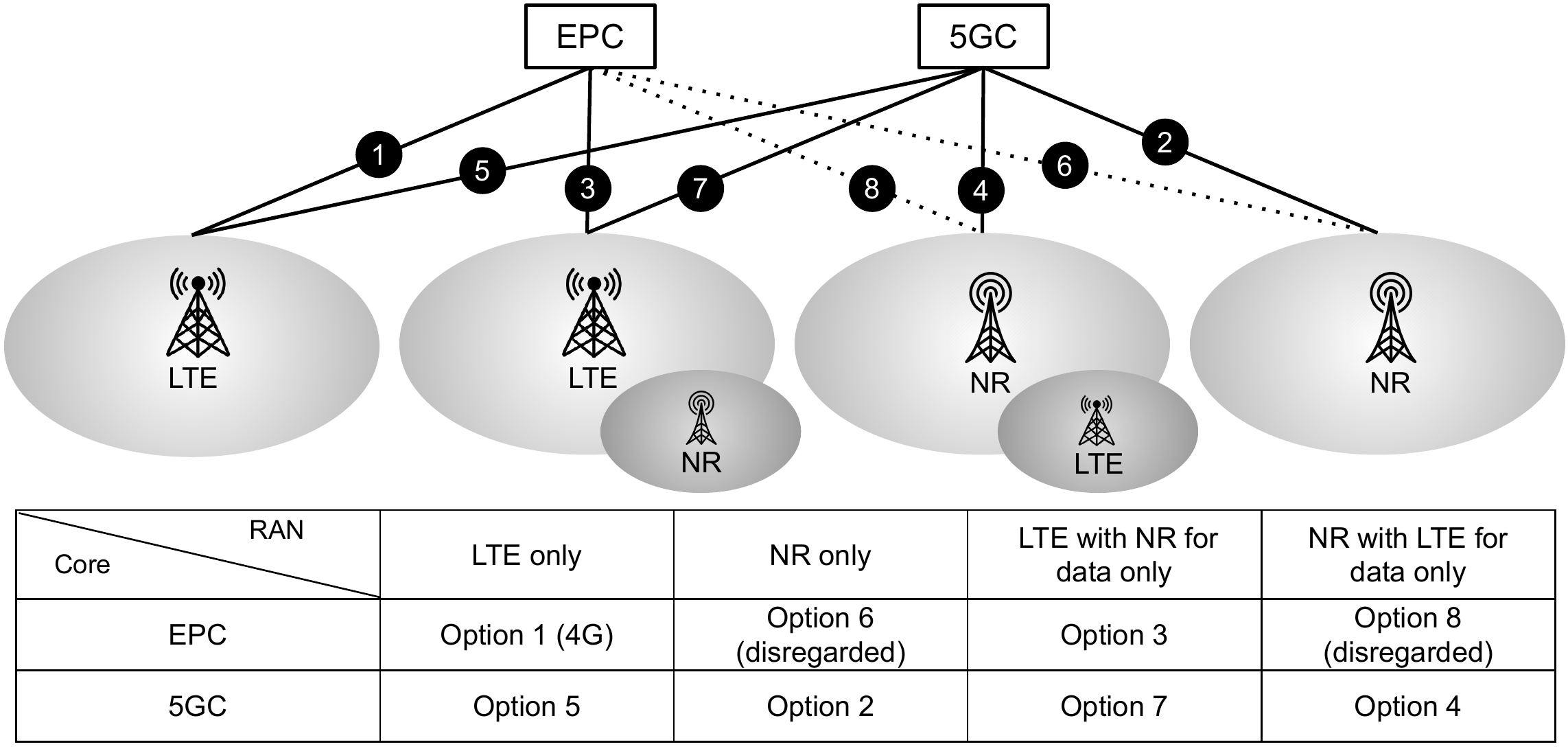}
  \end{center}
      \caption{Options defined during the initial phase of the 5G system definition.}
 \label{fig:opcoes_arq}
 \end{figure}

The 3GPP designed the 5G radio access network called Next Generation Radio Access Network (NG-RAN) and the core with a Service-Based Architecture (SBA) independently and interoperable. This is a different approach from the previous generations of mobile networks, in which RAN and core were designed in a coupled manner. Therefore, as of 5G, the standard provides for the integration of elements from different generations in different configurations. Specifically considering the transition between 4G and 5G, during the initial phase of the definition of the 5G architecture, a terminology was defined to identify the different variants that emerged from the integration between the access network and the core. The set of options are illustrated in Fig.~\ref{fig:opcoes_arq}.

  \begin{figure*}[htb]  
  \begin{center}
    \includegraphics[width=0.8\textwidth]{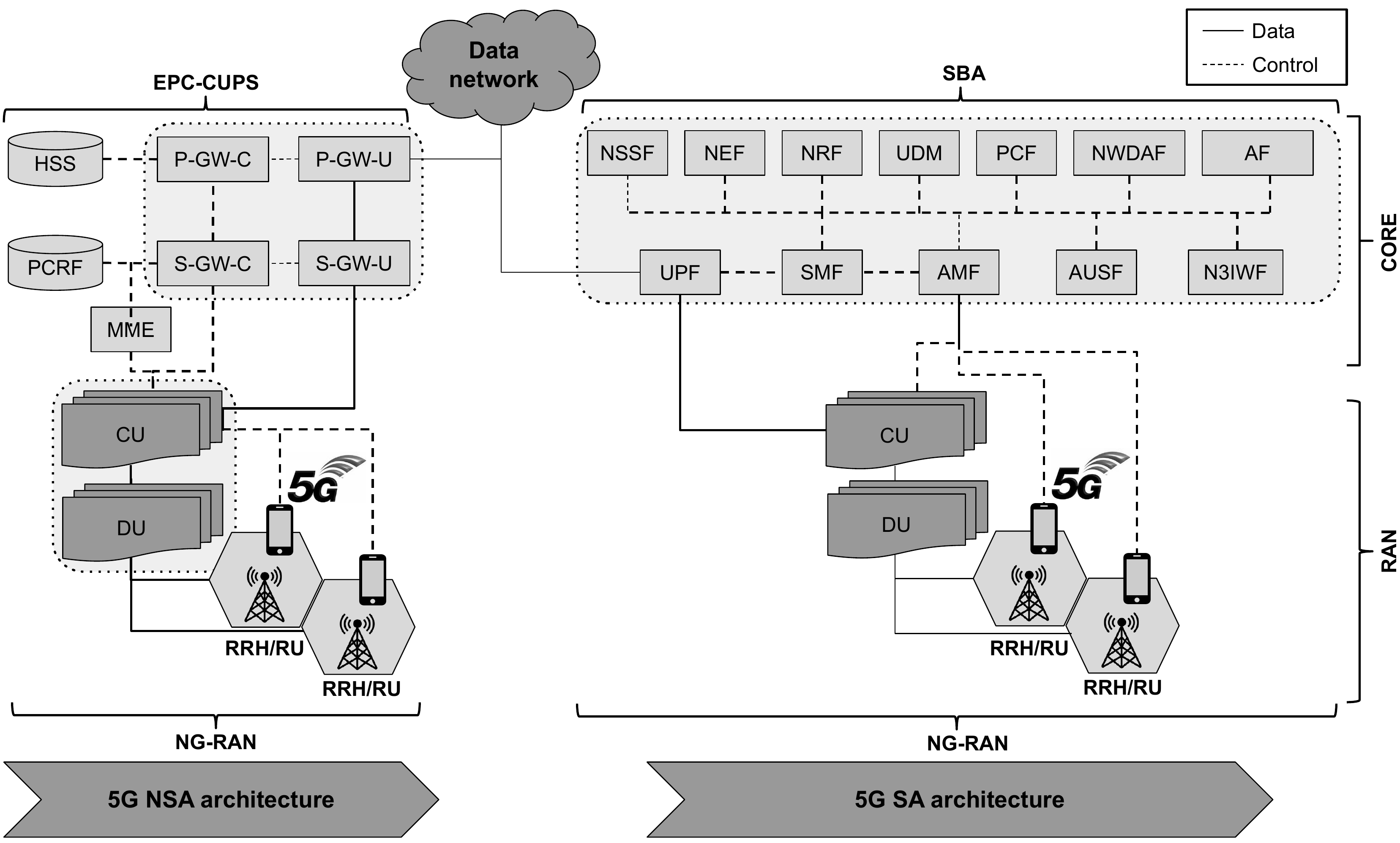}
   \end{center}
    \caption{5G architectures: Non-Stand-Alone (left) and Stand-Alone (right).}
 \label{fig:arch5g}
 \end{figure*}
In Fig. 6, option 1 corresponds to the existing 4G system and is standardized in previous releases of 3GPP. Options 6 and 8 were disregarded, as they would introduce many limitations to the 5G New Radio (NR) to provide backward compatibility with EPC. Although options 2, 3, 4, 5, and 7 were considered in the specification work, it was defined that the priorities would be options 2 and 3. Therefore, Release 15 of 3GPP introduced the Non-Stand Alone (NSA) architecture for option 3, and Stand-Alone (SA) architecture for option 2.

The NSA architecture was proposed in May 2018, and six months later, 3GPP presented the specifications for the SA architecture. NSA is a fast way to provide high data throughput and high connectivity, as it allows the use of existing network assets, without the need to implement a new complete end-to-end system for the 5G network, \textit{i.e.}, NSA is a transition architecture from 4G to 5G. In the NSA, only the radio technology needs to be updated. However, to better exploit 5G's capacity, aiming at new services, a modern architecture independent of the existing 4G system is necessary. SA is considered the ultimate 5G architecture, including improvements in radio transmission and a native 5G cloud core. The two architectures proposed in 3GPP Release 15 are illustrated in Fig.~\ref{fig:arch5g} and are discussed in the following subsections.

\subsection{5G Non-Stand Alone (NSA)}\label{sec:NSA}

The NSA architecture, also known as EN-DC (E-UTRAN-NR Dual Connectivity), meets the telecommunications industry's philosophy of introducing new technologies in the fastest and least disruptive manner. Although the motivations are clear, it is essential to highlight some disadvantages of NSA architecture. In this architecture, NR can be implemented only where 4G/LTE coverage already exists, indicating the absence of autonomy in the NSA. Furthermore, the available network functionality is limited to what is offered by LTE/EPC. Therefore, several novelties introduced in the 5G system are not available, for instance, network slicing, QoS management, flexibility in the edge computing, and the general extensibility of the 5G core, including new applications in an environment similar to the traditional cloud.

\begin{figure}[htb]  
  \begin{center}
 \includegraphics[width=0.5\textwidth]{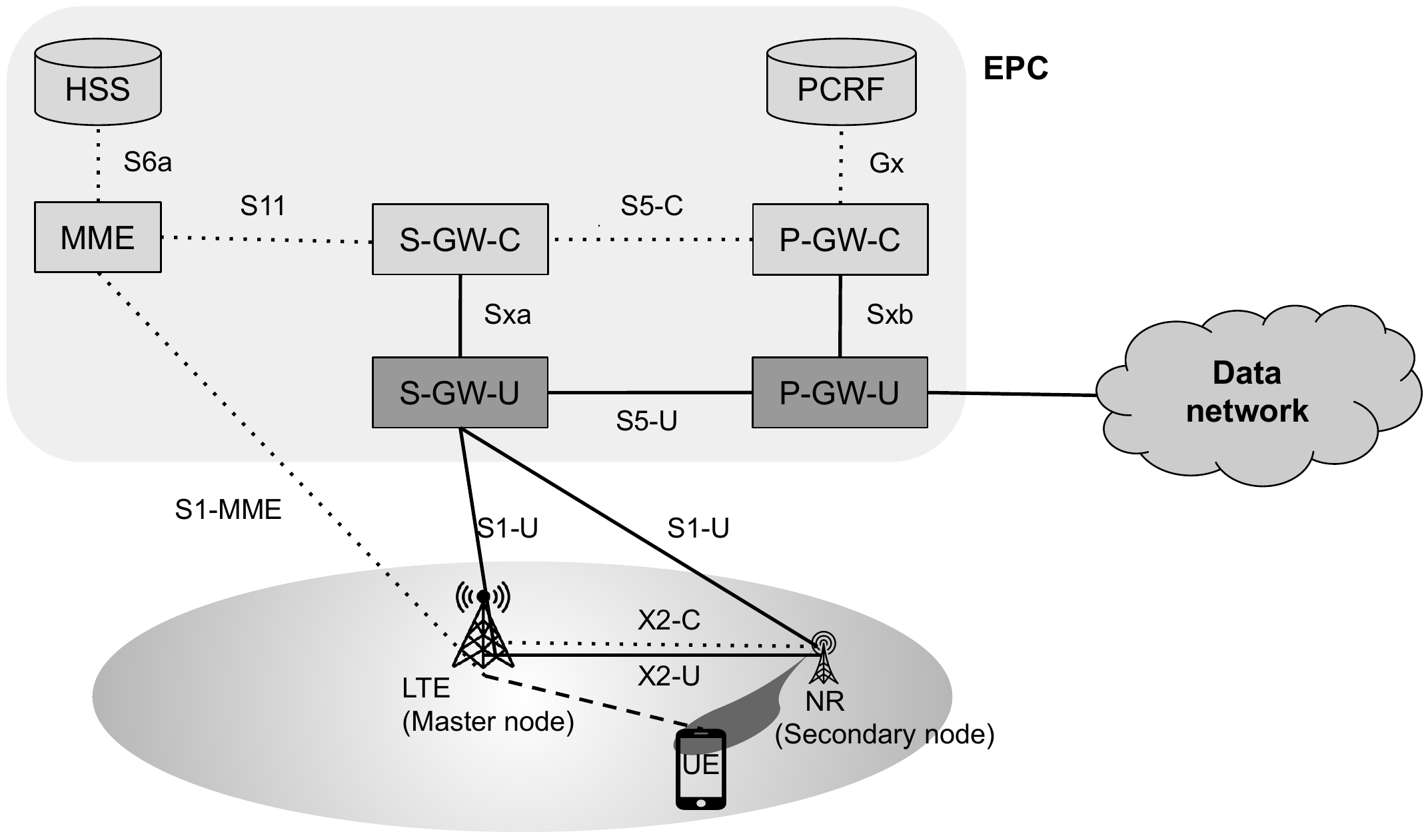}
   \end{center}
 \caption{NSA architecture, with emphasis on the selection between control and data in GWs and double connection (LTE and NR).}
 \label{fig:arq_NSA}
 \end{figure}

The NSA architectural design focuses mainly on allowing at least two 5G innovations to be introduced: (\textit{i}) noticeable increase in bandwidth capacity and network throughput, and (\textit{ii}) greater flexibility in the functions of the user plane provided by the gateways (S-GW and P-GW) of the EPC core. As illustrated in Fig.~\ref{fig:arq_NSA}, a separation between the control and user (or data) planes in the gateways, and a dual connection through LTE and NR are two significant features introduced in the NSA architecture. Three sets of technologies were introduced to implement these characteristics: (\textit{i}) Dedicated Core networks (DECOR) and enhanced DECOR, (\textit{ii}) CUPS and (\textit {iii}) NR as secondary Radio Access Technology (RAT). In the following, these technologies are briefly described.

\subsection*{Dedicated Core networks (DECOR) e enhanced DECOR}

\begin{figure}[htb]
\centering
    \subfigure[pre-DECOR]
    {\label{subfig:pre-decor}
    \includegraphics[width=.44\textwidth]{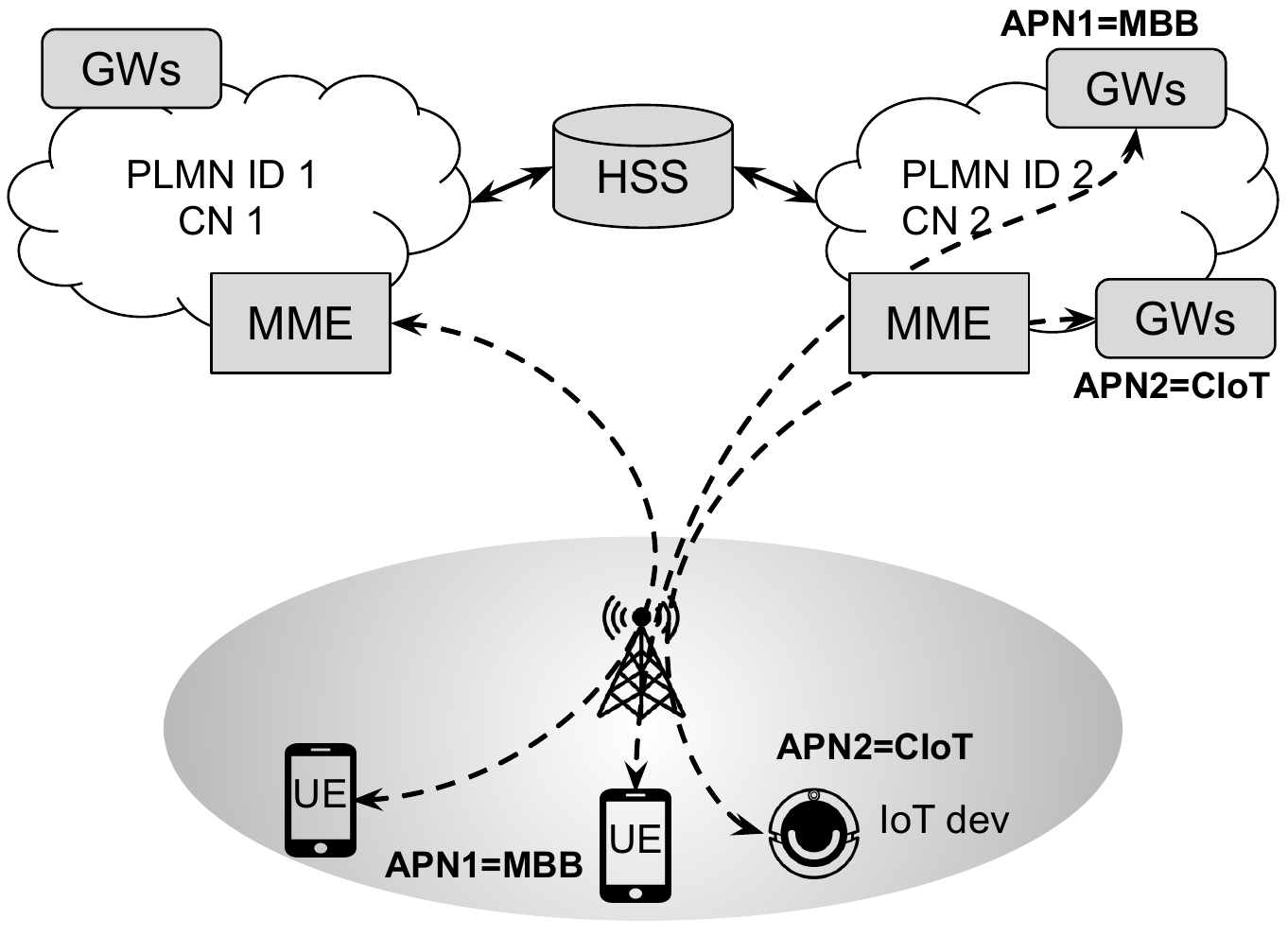}}
    \hfil
    \subfigure[DECOR and eDECOR]
    {\label{subfig:e-decor}
    \includegraphics[width=.44\textwidth]{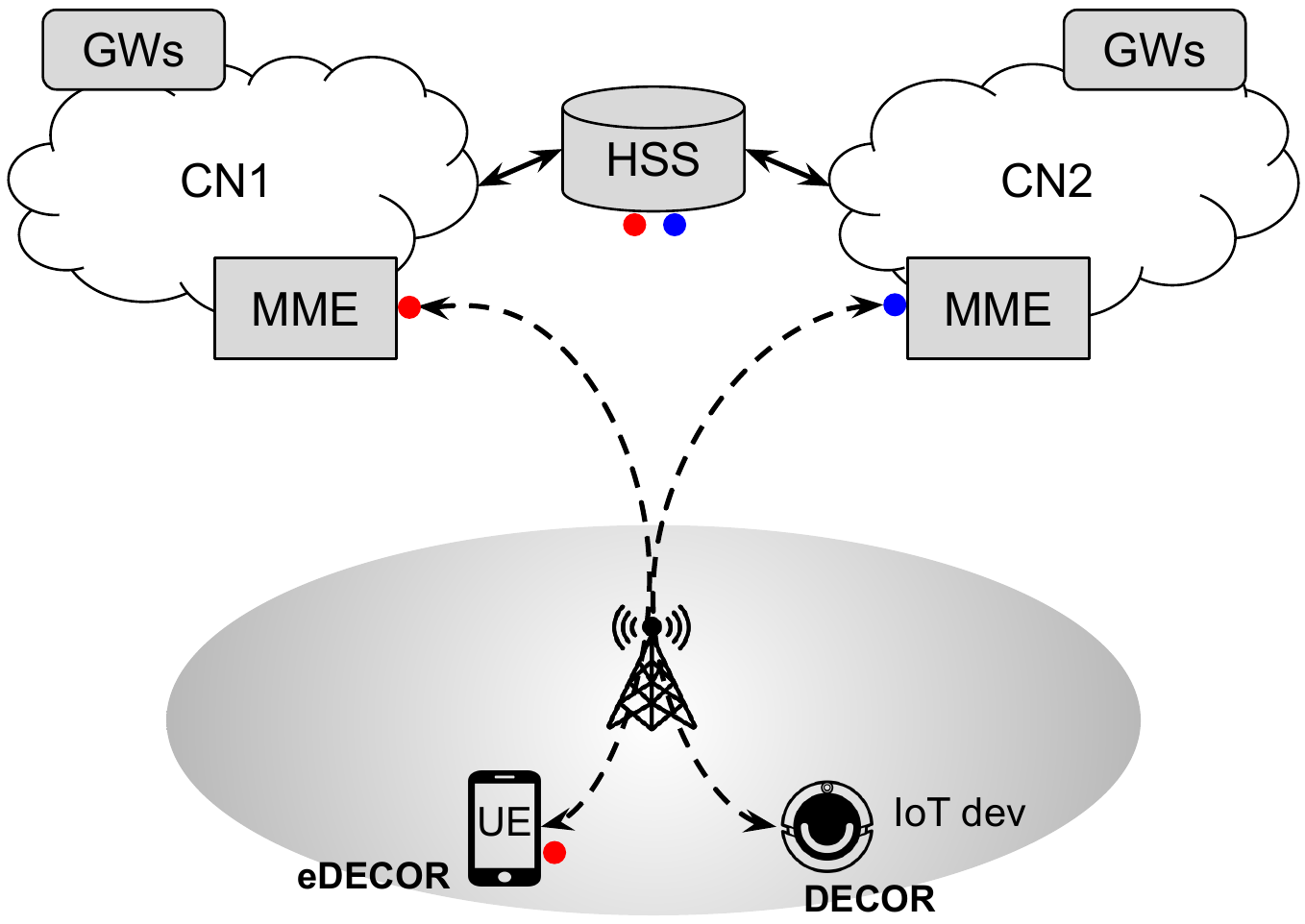}}
    \caption{Example of a pre-DECOR configuration, using APNs, compared to using DECOR and eDECOR.}
\label{fig:decor}
\end{figure}

A mobile network operator (MNO) is recognized by its Public Land Mobile Network (PLMN) identifier, which corresponds to a mobile network core. However, even in 4G, there was a significant demand to make this approach more flexible and allow a single operator to instantiate multiple cores and direct users to the appropriate core, according to the required service. Before introducing (e)DECOR, the separation of cores was possible using different PLMN identifiers, i.e., instantiating independent core networks. An alternative solution was to use different Access Point Names (APNs) to direct users to various service networks, which were associated with multiple user plane entities, in particular, P-GWs. Both approaches, illustrated in Fig.~\ref{subfig:pre-decor}, are not flexible.

Through the introduction of (e)DECOR, an operator can deploy multiple Dedicated Core Networks (DCNs) within a PLMN with each DCN consisting of one or more nodes of the core network (\textit{e.g.}, only MME, MME with GWs, MME, GWs, and PCRF). Each DCN can be dedicated to serving a different type of UE, separating certain types of traffic on specific core network nodes and, if necessary, adjusting them differently from the rest of the nodes of the core network. With DECOR, the information needed to identify how to route user traffic is obtained by MME by accessing only the HSS. On the other hand, eDECOR requires UE to provide specific information (\textit{i.e.}, the preferred DCN) to facilitate fast and optimal selection of the dedicated core network. Fig.~\ref{subfig:e-decor} illustrates the use of DECOR (with information initially only in the HSS, represented by the blue circle) and eDECOR (with information also in the UE, represented by the red circles) to access different DCNs.

The (e)DECOR can be considered the forerunner of network slicing, already introducing some characteristics similar to the QoS differentiation and the creation of multiple instances of some of the core components to serve different services. On the other hand, network slicing is a more advanced concept that allows the mobile network operator to perform end-to-end slicing, from the radio resources, through the entire access network and including the whole core. Network slicing should allow the creation of resource slices that are partially or totally isolated, physically, or logically.

\subsection*{Control and User Plane Separation (CUPS)}

\begin{figure}[htb]  
  \begin{center}
 \includegraphics[width=0.5\textwidth]{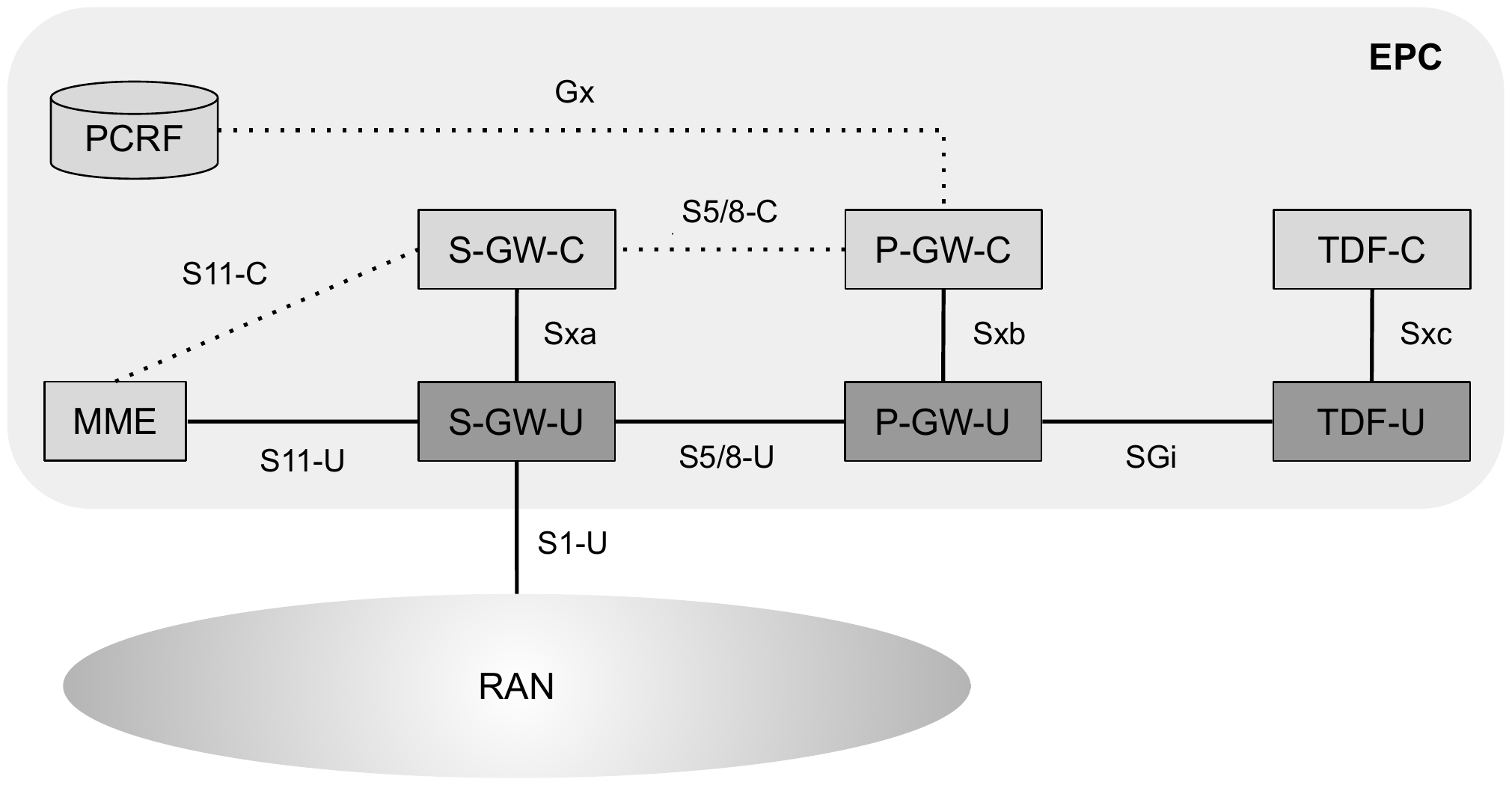}
   \end{center}
 \caption{Basic EPC architecture with the separation between the control and user planes.}
 \label{fig:cups}
 \end{figure}

The separation between control and user plans emerged from the need to independently dimension the user plane and control functions in the core network for session management and data services. Before the introduction of CUPS, it was not possible to deploy GW components with only the user (data) function or to independently dimension, in a standardized way, the parts of the control plane and user plane. The need for this separation became very clear as operators began to consider the impacts of third-party on their internal resources, such as (narrowband) IoT, Mobile Broadband (MBB), and also the growth of Internet-based Over-The-Top (OTT) services, such as streaming video, content sharing, and social media communication. As shown in Fig.~\ref{fig:cups}, CUPS can be applied to the following elements of the EPS: S-GW, P-GW and TDF (Traffic Detection Function).

As shown in Fig.~\ref{fig:cups}, the control plane and the user plane of each element (S-GW, P-GW, and TDF) use an Sx interface (a, b and c, respectively) to carry out the necessary communication. The Sx interface offers procedures for establishing, modifying, and closing, thus, providing the support for the control plane and user plane operations between the components of each separated node. 3GPP standardized the Packet Forwarding Control Protocol (PFCP) to support these functionalities in Sx, and this protocol was also adopted in 5G systems.

Taking S-GW and P-GW as examples, it is interesting to highlight some design features of CUPS. It was evident that not all scenarios would require separation between the planes (control and user) and that the most common deployment scenarios would have to contemplate the coexistence of nodes with and without separation in a single network. Therefore, it was defined that the separation between the planes should not have any impact on other components of the core, such as MME, PCRF, billing system, and subscription management system. CUPS was also designed not to impact any procedure or protocol involving UEs and RAN elements. Consequently, other components of the network are unaware of whether S-GW and P-GW with which they interact have separate control and user planes or not.

\subsection*{NR as secondary access radio technology}

The dual connectivity (DC) of the NSA architecture allows UE and RAN to receive and transmit data through two base stations simultaneously. Thus, the DC provides the ability to use radio resources provided by two cells operating independently but connected to a single core. In the NSA context, the original motivation for dual connectivity is to increase user throughput, but it is also possible to provide greater mobility robustness and support load balancing between RAN nodes. Initially, the concept of dual connectivity was introduced for EPS with two groups of cells, both providing E-UTRAN resources. The BS used by the UE for initial connection (and also for all signaling with the core) is known as the master node and has the location where UE is associated. When UE reaches the connected state, the master node can request another BS (\textit{i.e.}, the secondary node) to offload data traffic. Later, the solution evolved to support multiple radio technologies (MR-DC - Multi-Radio DC), with EN-DC being the alternative that combines the technologies E-UTRAN and NR.

The requirement for spectrum flexibility was a determining factor for the adoption of OFDM-based technologies in LTE for 4G. It remains an important factor for planning and deploying NR to 5G. Throughout 3G and 4G, the need emerged for allocations in different spectrum frequency bands, several operating bandwidths, multiple duplexing schemes, and multiple-access schemes. However, the width and diversity of the spectrum used in NR represent one of the most peculiar characteristics of 5G. NR supports a large and diverse spectrum from 410~MHz to 52.6~GHz (and up to 100~GHz in future releases). Moreover, NR can use large bandwidths from 5~MHz to 3.2~GHz, supporting Time Division Duplex (TDD) and Frequency Division Duplex (FDD), as well as additional carriers for downlink (Supplementary Downlink -- SDL), or uplink (Supplementary Uplink -- SUL). In this context, 3GPP defined in the Release 15 two frequency ranges (FR): (\textit{i}) FR1 (410~MHz -- 7125~MHz) and (\textit{ii}) FR2 (24250~MHz -- 52600~MHz). FR2, also known as the millimeter-wave band, is the one that offers the greatest capacity but has different physical characteristics from FR1, presenting a challenge for deployment and use in 5G. In the following, we present some additional information about millimeter-waves, which are still a subject of significant academic interest~\cite{rangan:14, wang:18}.

In addition to being a spectral band with a low occupancy rate and less crowded, the millimeter-wave band can allow communications with rates on the order of gigabits per second, and it is quite attractive for mobile networks. However, these waves cause significant challenges in the design, implementation, and operation of these systems, as they suffer more considerable degradation and attenuation in the propagation. Moreover, these waves are more susceptible to physical blockages and atmospheric absorption. On the other hand, with the development of increasingly smaller antenna elements, thanks to its shorter wavelengths, the construction of antenna sets for the use of massive MIMO and beamforming is enabled, which helps in dealing with the propagation problems and improves significantly the frequency reuse and the spectral efficiency. However, beam forming/maintenance takes a considerable amount of time, and its improvement has been the subject of several studies~\cite{giordani:19, zhang:19}. Additionally, the problem becomes particularly challenging when considering the mobility of the users.

\subsection{5G Stand Alone (SA)}\label{sec:SA}

In the SA architecture, EPC is replaced by SBA (described in detail in Section~\ref{sec:core}), allowing the use of a set of virtual network functions that help other functions in the architecture or even for final users applications. In this way, an SA architecture consolidates the concept of decoupling of the data and control planes, allowing flexible and stateless positioning of virtual environments in the different network segments that make up a 5G system. For example, these segments can be seen as Edge, Fog, and Cloud, or even specifically in the RAN, we can find organized like Fronthaul, Midhaul, and Backhaul.

A significant evolution in the SA architecture is the design based on virtual network functions according to cloud's native model. This approach refers to how the functions of virtual networks are created and deployed flexibly, widely using the concept of cloud computing to develop, deploy, and manage services. For example, within this approach, the concept of microservices can be used for the implementation and expansion of a 5G system, in line with the fast evolution of information technology. One of the objectives of using microservices is to be able to decompose the components into functions used, with low granularity, to make the service light and with a high capacity for sharing. This objective perfectly fits the need to define eMBB, mMTC, and URLLC communication scenarios since it offers modularity, reuse of functions, and interoperability with heterogeneous networks. Furthermore, the introduction of microservices and interfaces to access them simplifies the processes of updating and maintaining 5G systems, reducing the cost of operation, and accelerating the introduction of new services.

\begin{figure}[htb]  
  \begin{center}
 \includegraphics[width=0.5\textwidth]{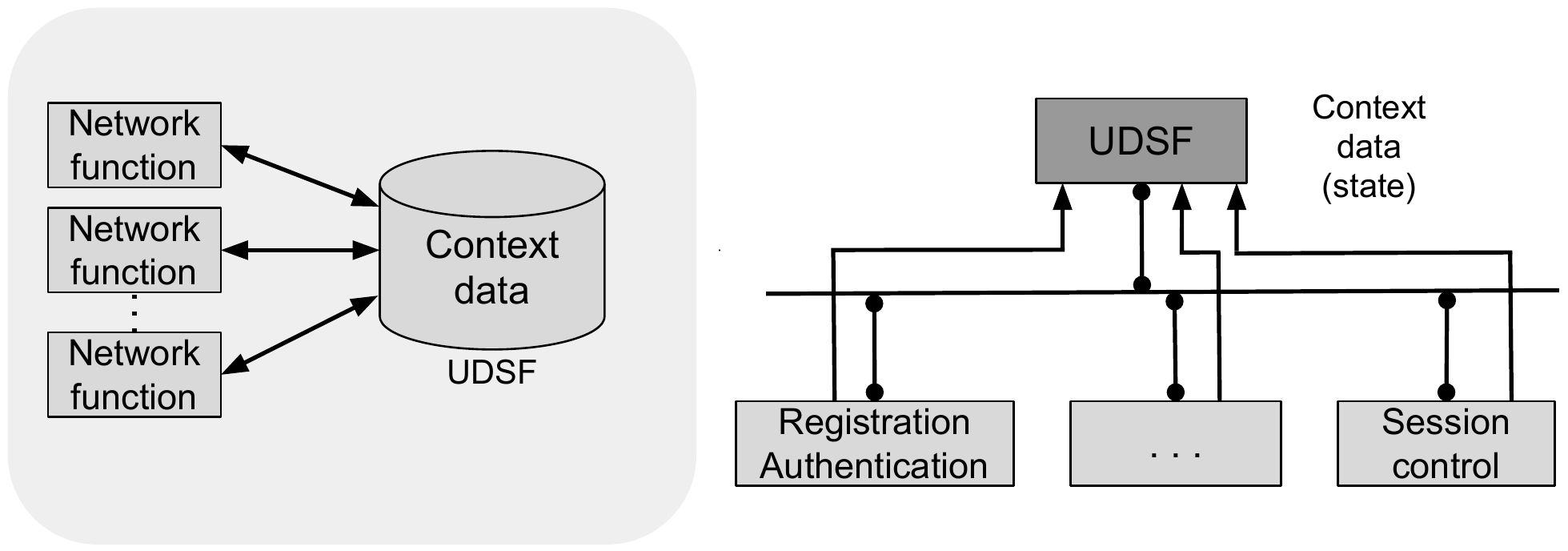}
   \end{center}
 \caption{Context data function of SA architecture.}
 \label{fig:stateless}
 \end{figure}

The full implementation of SA architecture presents some challenges. For example, historically, the mobile network generations are strongly affected in the context of UE sessions, featuring a state-based architecture. On the other hand, systems based on the native cloud computing model have a fundamental feature, stateless architecture, where the contexts of their states are saved in the databases. For instance, the SBA core has introduced an unstructured data storage function (UDFS) to control the setting of network functions, as can be seen in Fig.~\ref{fig:stateless}. This characteristic allows the system to enjoy benefits such as elasticity and dynamic management of microservices on mobile networks. One of the great challenges of 5G is to design and develop software components originally based in states, through the stack of protocols used by 3GPP, for stateless microservices, enjoying the benefits of using the native cloud computing.

\begin{figure}[htb]  
  \begin{center}
 \includegraphics[width=0.5\textwidth]{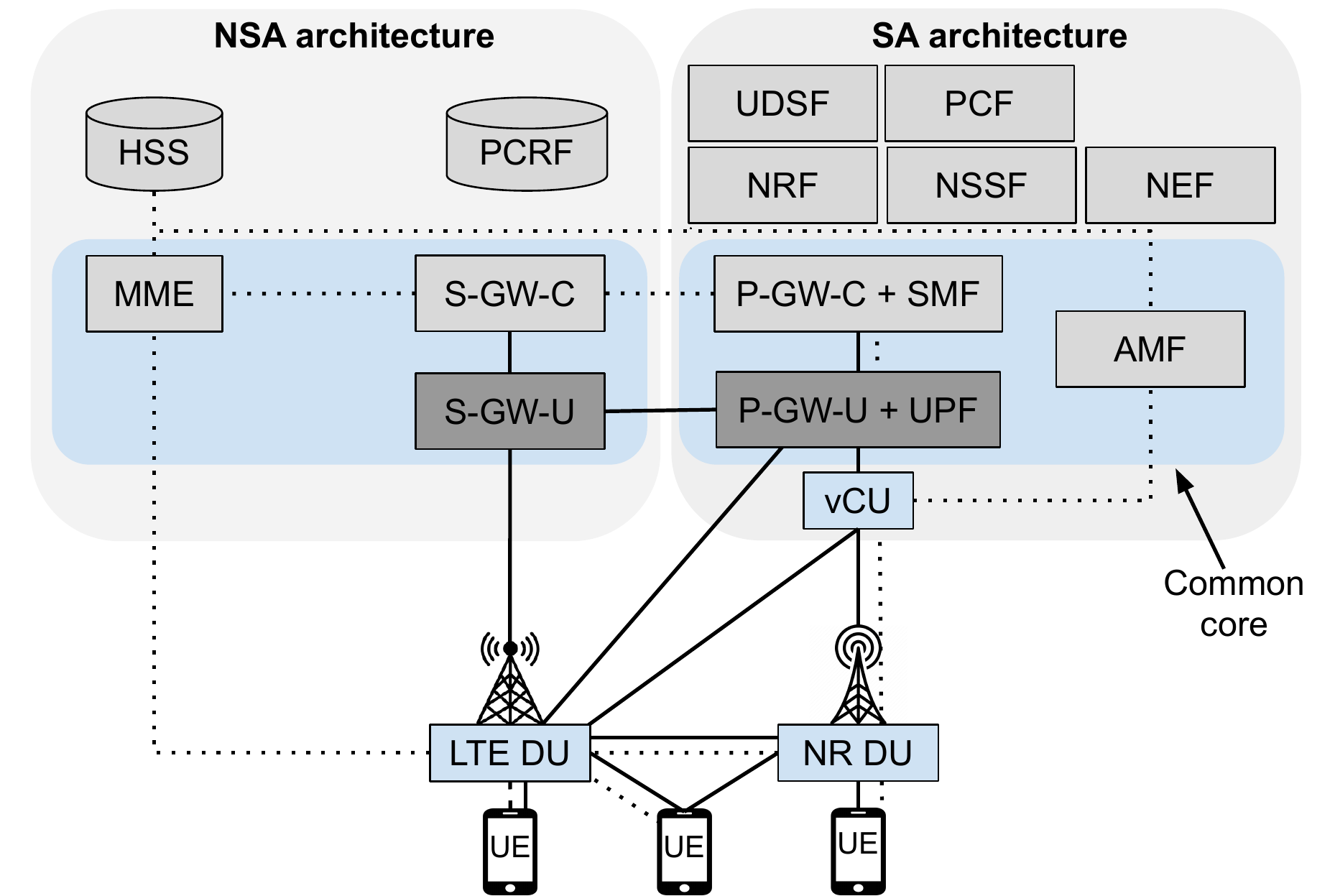}
   \end{center}
 \caption{Proposal for the coexistence of NSA and SA architectures.}
 \label{fig:NSAeSA}
\end{figure}

In the SA architecture, the fundamental characteristic is that the elements are defined as network functions, which provide services to other functions, or any other authorized \textit{consumer}, through programming interfaces. In this way, the architecture provides extensive modularization and reusability in complete isolation between DP and CP. Due to this high modularization, the eventual migration of the NSA architecture to SA must be hidden to the end-user. Moreover, it is worth mentioning that NSA and SA are not competing for each other, but rather an evolutionary path for adopting the innovations introduced by 5G. The intention is to start with the NSA and gradually migrate to SA over time, especially for operators that already have significant investment in 4G networks. For some time, NSA and SA must coexist, and several approaches are possible. Fig.~\ref{fig:NSAeSA} illustrates a potential solution proposed by Samsung~\cite{samsung-2019}, in which an intermediate architecture emerges, using a common core.\\
\\
\textbf{Open-source software initiatives in 5G}\\
The OpenAirInterface Software Alliance also has initiated the development of software for 5G\footnote{\url{https://gitlab.eurecom.fr/oai/openairinterface5g}} involving both RAN and core. Nowadays, only basic features are available~\cite{kaltenberger:19}, but it is possible to observe the evolution of the new gNB and the 5G-NR. An operational 5G core from the OpenAirInterface is still not available, but there are efforts in several components, such as AMF, SMF, UDM, AUSF, and UPF\footnote{\url{https://www.openairinterface.org/docs/workshop/8_Fall2019Workshop-Beijing/Training/2019-12-03-NGUYEN-DU.pdf}}. Another initiative is the free5GC\footnote{https://www.free5gc.org/}, an open-source project focused on implementing the 5G core network (5GC) defined in 3GPP Release 15 (R15) and beyond. The code is organized in stages, in which stage 1 is actually an EPC implementation improved with AMF, SMF and UPF. Stages 2 and 3 are `pure' 5G core implementations, in which the latter is the most functional one. Yet another initiative is the Open Core Network group\footnote{\url{https://telecominfraproject.com/open-core-network/}} that still does not have any software available, but that is supported by the Telecom Infra Project (TIP). TIP has been involved in several other open-source initiatives related to the telecommunications industry.

\subsection{Beyond 5G}

Although many countries, such as Brazil, are still discussing and preparing for the deployment of 5G networks, 2019 is the year when effective adoption began in several countries. Germany, China, South Korea, United States, Finland, and the United Kingdom are examples of countries that are already deploying, testing, and effectively using 5G networks. As expected, the scientific community has already begun to investigate questions 1) about the limits of 5G networks, 2) about the (new or existing) applications that would benefit from wireless communication, but would be supported by 5G networks, 3) about the technologies that need to evolve or be created to meet new potential demands, among several other themes~\cite{saad:19, giordani:20}. Although it is not possible to predict precisely what the next generation of mobile wireless networks will be and which applications will justify its adoption, but some topics are recurrent in the initial discussions and investigations and are likely to be part of the evolution of 5G networks. Among these themes are related with the efficient allocation of diverse resources in wireless networks (\textit{e.g.}, sub-6GHz bands, millimeter-waves and, in the future, bands in THz \cite{chen2019survey}), wide use of machine learning and artificial intelligence, and applications that are not served by 5G networks, such as multisensory extended reality and wireless brain-computer interactions.

5G networks have made the need to consider multiple types of wireless resources notable, in addition to the traditional licensed sub-3GHz bands. In 5G networks, it is possible to use the 3.5GHz band, unlicensed 5GHz band, millimeter-waves, and D2D communications. Additionally, there are already Multi-access Edge Computing (MEC) resources that allow improving the communication of the devices through various strategies such as caching, downloading of processing, use of context information, allocation of bandwidth, among others. However, a reliable high-band offer using millimeter-waves (later from the THz band) in mobile scenarios is a challenge that will hardly be overcome on a large scale in 5G networks. Due to the high directionality of the millimeter-wave beams, the alignment between the antennas of mobile devices and BSs must be remade with high regularity. Currently, this means disruptions in communication, but as the number of BSs (both millimeter-waves and sub-6GHz) grows, it is possible to mitigate the problem through the appropriate resources allocation policies. However, this allocation is challenging, as it involves multiple devices, multiple BSs, different resources, and must be performed dynamically in short time scales to meet the mobility of users~\cite{semiari:19}. Based on this context, there is a large amount of data that needs to be processed quickly for decision-making regarding resource allocation.

In addition to optimization models, the context described can also make extensive use of machine learning and artificial intelligence techniques. These techniques were initially explored at 4G through the concept of Self-Organizing Networks (SON)~\cite{peng:13}. In 5G networks, machine learning and artificial intelligence should have broader use, given the large volume of data to be manipulated and the greater complexity of the network~\cite{saad:19}. However, widespread adoption is expected only after the consolidation of 5G networks and in the next generation of mobile communications~\cite{letaief:19}. Moreover, the increase in the volume of data and the complexity of the infrastructure and recent evolution in machine learning and intelligence techniques have motivated its more extensive use. However, for artificial intelligence-based solutions to be widely used in communication networks, they must be built from the perspective of Explainable Artificial Intelligence (XIA)~\cite{adadi:18}. Understanding the behavior of solutions based on artificial intelligence is an essential requirement for its adoption in communication infrastructures.

5G networks are introducing new communication technologies to serve a wide range of applications and services. For example, eMBB, URLLC, and mMTC are part of the 5G system to support applications such as 4K and 8K videos, autonomous cars, virtual and augmented reality, and Industry 4.0. However, applications are already emerging that are not adequately served by 5G networks, such as multisensory extended reality, holographic telepresence, wireless brain-computer interactions, and connected autonomous and robotic systems~\cite{saad:19, giordani:20}. In general, these applications have new QoS requirements that can motivate the introduction of new classes of service such as Mobile Broadband
Reliable Low Latency Communication (MBRLLC) and Massive Ultra-Reliable Low Latency Communications (mURLLC -- massive URLLC). On the other hand, to exploit the available resources more efficiently, it is vital to investigate and adequately characterize the demand for applications, taking into account QoS requirements, Quality of Experience (QoE), and quality of Physical Experience (QoPE)~\cite{saad:19, taleb:19}.
\section{The Radio Access Network}\label{sec:RAN}

\begin{figure*}[ht]
 \begin{center}
\includegraphics[width=0.8\textwidth]{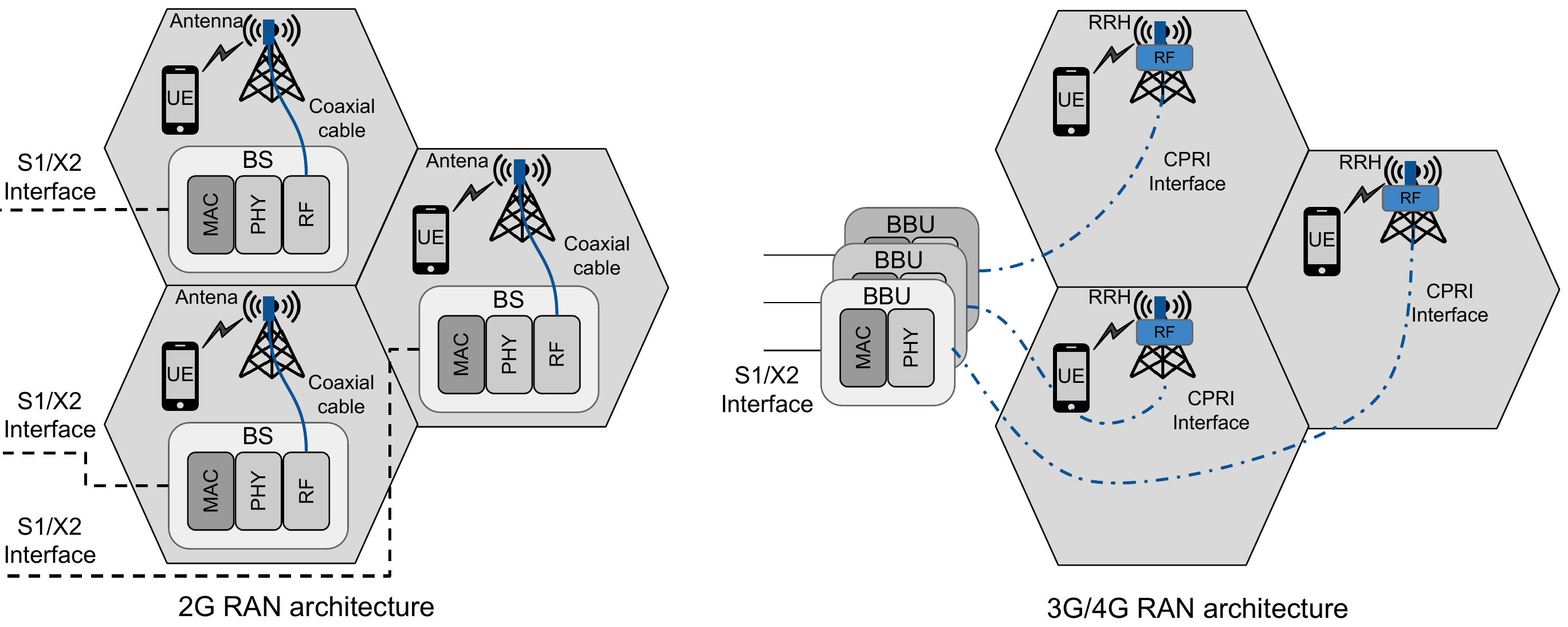}
  \end{center}
\caption{Evolution of RAN architectures.}
\label{fig:evolucaoRAN}
\end{figure*}

A RAN is responsible for managing the air interface to keep a large number of users connected. Therefore, RAN needs to efficiently manage the RF spectrum, performing a series of tasks, such as radio resource management, connection mobility control, dynamic allocation of resources to users' equipment, compression and security in the physical layer, session management, and QoS flow, among other functions.

Initially, on a traditional RAN, the RF front-end and baseband processing functionalities were integrated within a BS. The antenna module was located a few meters from the radio module, connected with a coaxial cable, and showed high transmission losses. This architecture was noted on 1G and 2G mobile networks. In recent years, the RF front-end has been decoupled from the baseband processing module, allowing mobile operators to replace this module according to user demand. This architectural decoupling allowed advances in centralization and virtualization in the baseband processing module and the RF front-end. Next, Subsection \ref{subsec:CRAN} presents the evolution of the concept of centralization of baseband architecture, and Subsection \ref{subsec:VRAN} presents the virtualization of RAN, which started in 4G and continues to evolve in networks 5G.

\subsection{RAN centralization} \label{subsec:CRAN}

The high data transmission rate between the digital baseband processing domain and the analog domain of the RF front-end with the antenna requires a high bandwidth bus that connects these two domains. For many years, this requirement has limited the BS design to specialized hardware components with that high-performance bus. This limitation was overcome with the use of fiber optics on this bus, reducing data loss and increasing the distance between the connected domains. In 3G/4G generations, especially since Release 8 of the 3GPP, the baseband processing started to be implemented in BBUs, \textit{i.e.}, dedicated and specialized hardware that implements a RAT. At the same time, RRH integrates the RF front-end with the antenna. In Fig.~\ref{fig:evolucaoRAN}, the evolution of the RAN architecture from 2G to 3G/4G is illustrated.

In 3G/4G, the RF front-end and the baseband processing are separated, as can be seen on the right side of Fig.~\ref{fig:evolucaoRAN}. RRH also called a Radio Unit (RU), has a fiber optic interface and performs analog/digital conversion and vice-versa, power amplification, and signal filtering. The baseband processing, now performed at BBU, is isolated and independent from RU. This architecture is considered decentralized since BBU performs its operation separately from RU. Recent advances have boosted the bandwidth of optical fibers allowing the incorporation of cloud-based architectures into the Cloud Radio Access Network (C-RAN). In this type of architecture, BBUs can be located in a data center (called centralized baseband architectures), allowing cost reduction through centralized maintenance and by the elasticity of cloud computing. In this new configuration, RRHs can be geographically separated from a set of BBUs by up to approximately 40~km. However, it is important to note that a BBU is limited to processing RRH signals within a maximum distance, determined according to the delay restrictions. This delay is mainly influenced by three factors: (\textit{i}) distance between BBU and RRH, (\textit{ii}) channel conditions, and (\textit{iii}) available processing capacity. According to Marotta et al. \cite{Marotta:18}, the processing capacity should be increased significantly for RRHs that are experiencing low Signal-to-Noise Ratio (SNR) and long distance from the BBU. 

The concepts involved in C-RAN have been drawn attention to the initial implementation of 5G (Architecture NSA described in Subsection~\ref{sec:NSA}), mainly considering the benefits of cost reduction and maintenance. However, BBUs are hardware-based platforms using specialized digital signal processors. Including a long-term objective, it is essential to replace BBUs based on specialized hardware with software using general-purpose hardware, \textit{i.e.}, Virtual BBUs (vBBUs). In the same way, we can think of a radio virtualization layer that allows several heterogeneous access technologies coexisting on the same RRH. This coexistence uses innovative baseband processing techniques to divide and abstract an RRH into multiple virtual RRHs (vRRHs).  This vision of RAN virtualization is aligned with the SA Architecture (described in Subsection~\ref{sec:SA}) proposed in Release 15 of 3GPP, where, for example, the use of vBBUs is an Network Functions Virtualization (NFV) use case to provide services on a 5G system.

\subsection{RAN virtualization} \label{subsec:VRAN}

The RAN virtualization (vRAN) has received prominence in 5G systems because it allows us to create, manage, and configure RANs dynamically, meeting specific requirements of each service. Furthermore, the vRAN concept opens up new business models in which service providers can rent vRANs from infrastructure providers. In this scenario, the infrastructure provider controls the entire physical resource, including the RF spectrum, the physical RRH, the hardware resources in the data centers (\textit{i.e.}, servers with processing, memory, and storage) and the physical network. A service provider can hire from one infrastructure provider one or more vRANs, including at least one virtual BS, \textit{i.e.}, a vRRH connected to a vBBU.

\begin{figure*}[htb]
 \begin{center}
\includegraphics[width=0.7\textwidth]{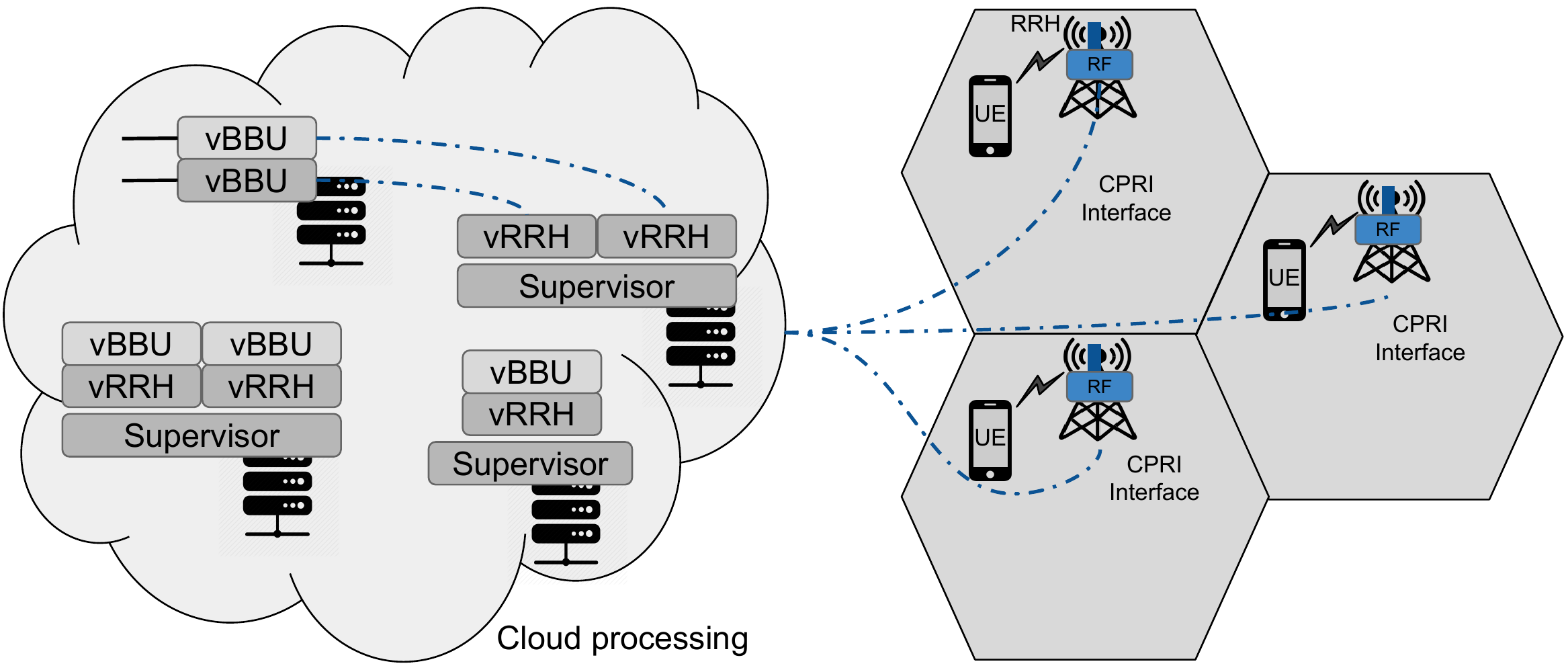}
  \end{center}
\caption{Virtualized RAN.}
\label{fig:vRAN}
\end{figure*}

Isolation, programmability, and adaptability are essential properties for customizing vRAN to accommodate the different services provided for 5G networks. For example, the combination of BBU and RRH slicing can instantiate an end-to-end vRAN over physical infrastructure. More specifically, slicing and virtualization can be implemented in a BBU to allow multiple vBBUs to run on the same physical hardware. Likewise, one RRH or a combination of them can support multiple vRRHs. These elements applied to vRAN, constitute the pillars to provide multi-services for future mobile networks.

A set of vBBU and vRRHs can be run on General Purpose Processor (GPPs), taking advantage of the highly optimized signal processing libraries and taking advantage of the ever-increasing evolution of processors, such as higher processing power and energy efficiency, such as can be seen in Fig. \ref{fig:vRAN}. Recently, the 3GPP RAN3 working group \cite{3gpp:TR38.801} considered dividing a vBBU into two new entities, named Data Unit (DU) and Control Unit (CU). DU can host time-limited physical layer functions, while CU hosts non-critical function resources, such as the MAC layer and higher control services. DU implementation is expected to cover an area of 10 to 20 km in radius, while CU implementation should cover areas from 100 to 200 km.

In Release 15 of 3GPP, the cloud concept was maintained in the RAN, but the name was changed to NG-RAN using an interface called NG. Due to interoperability between 4G and 5G, DU and CU have been renamed gNB-DU and gNB-CU, respectively. A gNB is responsible for some tasks, such as radio resource management, connection mobility control, dynamic allocation of resources to users' equipment, physical layer compression and security, session management, and QoS flow, among other functions. Therefore, the gNB protocol stack is detailed, considering PHY, MAC, RLC, PDCP, RRC, etc.

\begin{figure}[htb]
 \begin{center}
\includegraphics[width=0.45\textwidth]{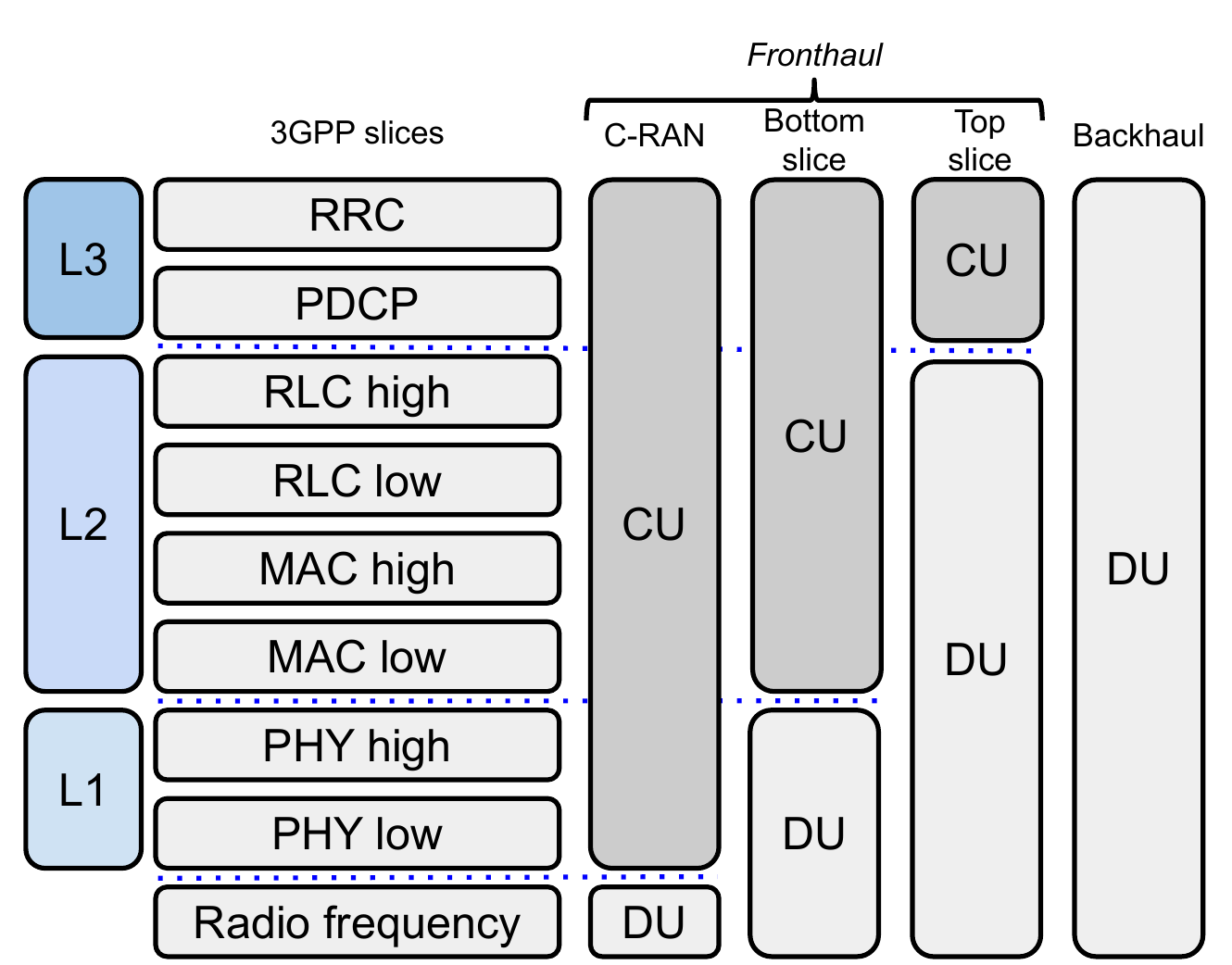}
  \end{center}
\caption{Division of functions between the central and the distributed units.}
\label{fig:splitsLayers}
\end{figure}

Virtualization also introduced the possibility of dividing the gNB protocol stack into network functions. In this context, 3GPP proposed eight options for the functional division between centralized and distributed units \cite{3gpp:TR38.801}, considering transport requirements, in particular flow and latency. González-Día et al. ~\cite{Gonzalez-Diaz:19} evaluated, in an experimental setting, three slicing options for fronthaul, in addition to an option for the DU backhaul, as illustrated in Fig.~\ref{fig:splitsLayers}. These eight slicing options proposed by 3GPP are still the subject of research ~\cite{fonseca-sbrc-2019} and development by academia and industry, as requirements for virtualization, processing, and functionality still need to be investigated.

It is important to note that this new flexible architecture of RAN composed of distributed data units (DU/gNB-DU) and control (CU/gNB-CU) brought changes to the transport network between the access and the core of the 5G system. Currently, the transport network is being redesigned and developed considering segments of fronthaul, midhaul, and backhaul. The fronthaul is responsible for the communication between RRH/vRRH and DU/gNB-DU. In the midhaul segment, communication takes place between DU/gNB-DU and CU/gNB-CU. Finally, the backhaul performs the communication between CU/gNB-CU and the new core of the 5G system. Additionally, approaches are expected to adopt the integration between transport segments, in a configuration of crosshaul~\cite {Gonzalez-Diaz:19,fonseca-sbrc-2019}, in which the objective is to explore the efficient use of resources high-cost transportation. Orchestrating the workloads of vRANs in this new architecture is a subject of considerable research today.

\subsection{RAN demonstration} \label{subsec:demo_RAN}

\subsection*{Goals}

One objective of the demonstration is to present practically the functionalities of a RAN, using open software and hardware. Another goal is to briefly comment on the software installation and configuration processes that implement the RAN, making available material for replication of this demonstration.

\begin{figure*}[htb] 
 \begin{center}
\includegraphics[width=0.9\textwidth]{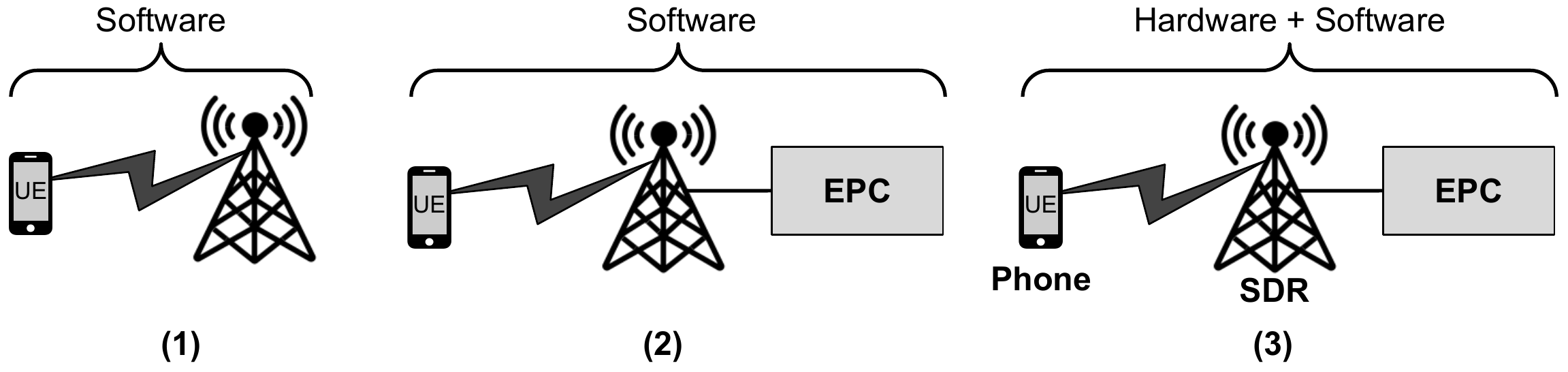}
  \end{center}
\caption{Experiments with a softwarized RAN.}
\label{fig:demo1}
\end{figure*}

\subsection*{Description}

In this demonstration, we create an operational eNodeB, \textit{i.e.}, the main element of a RAN, based on LTE technology and using open-source software. In addition to the RAN, the software is also capable of emulating functional UEs. The demonstration is organized in 3 experiments: (1) UE and RAN emulated by software, without the core; (2) UE, RAN and EPC core, all implemented in software; and (3) UE in hardware (conventional cell phone), RAN in hardware (SDR - Software-Defined Radio) and software, and EPC core implemented in software. All components are implemented using Docker containers that can be hosted on a cloud infrastructure. Fig.~\ref{fig:demo1} shows the software and hardware components that are used in the RAN demo.

\subsection*{Additional information}

During the tutorial, demonstration videos of the experiments are presented. Furthermore, manuals are available with details on how the tests can be replicated. Finally, the containers and any extra code produced by the authors needed to replicate the experiments are also publicly available.

The repository of this tutorial:\\
\url{https://github.com/LABORA-INF-UFG/NetSoft2020-Tutorial4}.
\section{Core network}
\label{sec:core}

\begin{figure}[hbt]  
 \begin{center}
\includegraphics[width=0.5\textwidth]{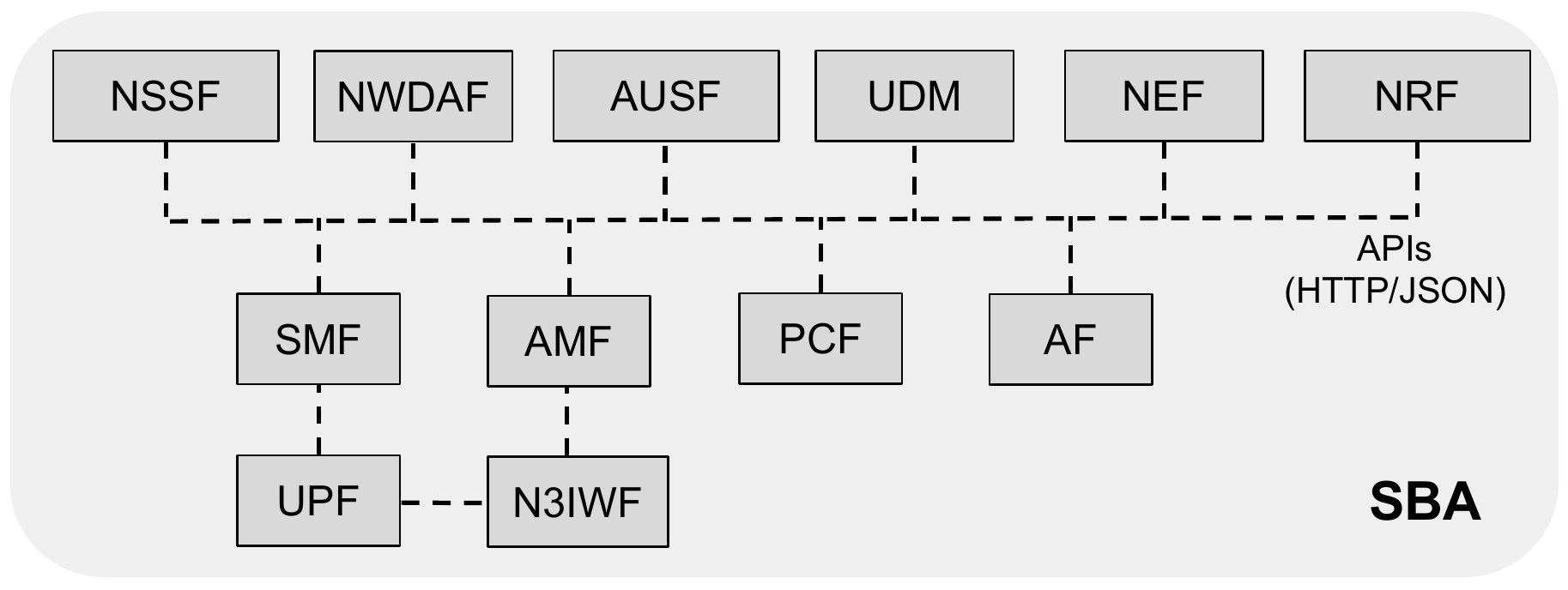}
  \end{center}
\caption{Core 5G software components.}
\label{5GCore}
\end{figure}

Based on the mobile network scope, the core can be characterized as the most critical element in the 5G system. 3GPP Release 15~\cite{3gpp:rel15nr21.915} defined the core as a set of components interconnected by a service layer. Each component has a specific set of responsibilities for consuming and providing services to/from other elements of the 5G system, through Application Programming Interface (API) defined by the standard. The structure of the software components and their interconnections via APIs is shown in Fig.~\ref{5GCore}. Release 15 introduced substantial changes in the way mobile networks are designed to support a wide range of services, each as distinct performance requirements~\cite{foukas2017network}.

Given the 5G core relevance, each of the elements that compose the core needs to be robust, resilient, highly available, and scalable. These characteristics are usually associated with the cloud computing concept. In the Release 15 context, each of the components that compose the new 5G core can be split into Virtual Network Functions (VNFs). In this way, each VNF can be available on-demand in a cloud infrastructure. The use based on the split of VNF is possible because physical interfaces like in previous architectures do not interconnect the components end-to-end. Instead, each of the components of the core must expose a set of software functionalities through SBA~\cite{karnouskos2012SOA}. In this context, each VNF offers one or more services for other VNFs. Considering the new definition of the SBA model, the exposure of functions applies only to the signaling context and not to the transfer of user data.

One of the communication methods used for an SBA is based on Representational State Transfer (REST) over HyperText Transfer Protocol (HTTP). This method consists of a set of rules and guidelines, widely used in the interconnection of distributed systems. The rules define how communication technologies on the Web access services through the use of APIs. The 3GPP expectation as this initiative is to make the task of extending the network resources simpler. Another aspect regarding the communication mechanism is that it should be seen as a logical option since the components specification for VNFs assumes that these components are running in a virtualized environment~\cite{imadali2018NFV}. 

The components of the 5G core can be seen as an interconnected network of services. Each VNF of a component performs specific responsibilities and interconnects with other VNFs producing and consuming services. The main characteristics of each of these components are discussed in the following subsections. We briefly describe the main components of the 5G core without the objective of listing all or making an exhaustive presentation. The complete list of components and their detailed description requires hundreds of pages of the standard or a whole book on it~\cite{hedman20195g}.

\subsection{Main Components}

In this subsection, we introduce the components needed to create a basic 5G core. In this way, we have named these components of main.

\subsection*{Access and Mobility Management Function (AMF)}

Mobility can be considered the essence of a 5G system. The mobility management begins when a new connection is established between a UE and the core of the network. This action triggers a set of procedures to identify UE, providing a security structure to offer a transport channel of messages. The main goal of the AMF component is to ensure that the communication process occurs cohesively and transparently, considering user mobility as a critical factor. Based on the functions implemented in AMF, the network can reach a specific user to notify about any messages or calls received, for example. Moreover, the AMF component can allow a particular UE to initiate a communication process with other UEs connected to RAN or with Internet access. Another fundamental functionality of AMF is to guarantee the connectivity holding the existing sessions when UEs move between different access points.

In 5G networks, there is a need to provide flexible support for a wide range of new users~\cite{foukas2017network}. Many of these users have specific requirements concerning mobility. For example, a particular UE used in a factory does not usually move, while UE in an autonomous vehicle or remotely controlled can present high mobility. To better support these different needs, 3GPP Release 15~\cite{3gpp:rel15nr21.915} splits the mobility procedures into three categories for the AMF component:

 \begin{itemize}
  \item Standard procedures can be characterized as a set of steps running when any UE requests a connection to the core. Among these steps, the security process is highlighted, composed of primary authentication, management of access keys, identification, and basic configuration of UE.
  \item Specific procedures have the function of managing the registration and periodic updating of the mobility of a given UE in AMF. Moreover, these procedures control the closing of the UE registration, provide scope for different access technologies. 
  \item Connection Management procedures are used to establish a secure communication process between a given UE and the core. Moreover, this procedure works when a specific UE needs to carry out a network resource reservation process for sending data.
\end{itemize}

Each of these categories of procedures aims to provide UEs with functionalities to establish connections with the core network, using the services associated with mobility.

\subsection*{Session Management Function (SMF)} 

The SMF component is responsible for managing the UE sessions, \textit{i.e.}, the sessions representing the connected users. The main responsibilities of SMF are activities of establishing, modifying, and releasing the individual sessions of UEs and allocating the IP address for each connected UE. However, the communication between UEs and SMF is performed indirectly through the AMF component. The function of AMF forwards the messages associated with the session of a given UE and the functions of the SMF component.

The SMF component's internal functions interact with VNFs of the other components through the producer/consumer model, defined regarding the SBA~\cite{karnouskos2012SOA}. For example, SMF is responsible for controlling different functions associated with the UPF component. This control includes the SMF's ability to configure the direction of data traffic associated with a UPF for a given UE session. Moreover, SMF must run monitoring and control actions on UPF. The SMF component interacts with functions related to PCF, aiming to execute the session policy of the UEs connected. This execution action can be described as one of the main tasks associated with a 5G system. For instance, this action determines the guidelines for data connectivity between a UE and the Data Network (DN).

\subsection*{User Plane Function (UPF)}

3GPP Release 15~\cite{3gpp:rel15nr21.915} introduces the UPF component as a fundamental function within the new SBA model. UPF can be seen as part of a split process between CP and UP, initially introduced in the Release 14~\cite{3gpp:rel14nr133.185} with CUPS. Decoupling data and control allows the SBA model to decentralize its components further. For example, it can direct activities such as packet processing to be placed closer to the edge network, increasing QoS for the user, and reducing network traffic.

SMF controls the functionalities implemented in the UPF component. The main function of UPF is associated with the forwarding and the data processing from UEs. This component is also responsible for generating notifications related to data traffic and the inspection of packets. UPF also works as a stable interconnection point between the core and any external networks. Moreover, UPF allows communication to happen transparently, hiding complex aspects associated with mobility. IP packets destined to a given UE are forwarded (from the Internet) to the respective UPF, which is attending that UE, even when UE is on the move. In general, the UPF component is responsible for:

\begin{itemize}
  \item Design the rules for the interconnection point between the core and any external networks;
  \item Serve as an external access point for Protocol Data Unit (PDU), interconnecting different data networks;
  \item Forward and route the packets and inspect these packets aiming to detect application characteristics;
  \item Apply the definitions associated with the management of the user data plane and provide information on the data traffic. 
\end{itemize}

\subsection*{Authentication Server Function (AUSF)}

AUSF is responsible for the authentication service of UEs through the access credentials provided by UDM. Moreover, AUSF provides services associated with the encryption for allowing secure information traffic and the execution of the roaming update processes, as well as other parameters related to UE.

Generally, the AUSF component provides services consumed by AMF functions, which request resources associated with the authentication process. For example, AUSF internally processes the requests of functions of the AMF component and after delegates to services provided by the UDM component to run the registry procedures in the data repository.

\subsection*{Unified Data Management (UDM)} 

The UDM component manages the users' data on the network in an only centralized element. UDM is similar to the HSS of the EPC core of the 4G networks. Through UDM, several VNFs of the SBA model can run many actions, such as registration and authentication of UEs, user identification, application of access rules and authorization, etc. The UDM component interacts directly with AMF, which forwards the requests from other components. Moreover, in scenarios where exist more than one instance of the AMF component on the network, UDM must control which instance is responsible for serving a given UE~\cite{toskala20205}. Among the UDM functionalities, we can highlight~\cite{etsiTS:123501}:

\begin{itemize}
    \item Generation of the Authentication and Key Agreement;
    \item Handling user identification;
    \item Support for privacy-protected signature identifier hiding;
    \item Access authorization based on the signature data (\textit{e.g.}, restrictions associated with mobility)
    \item Signature management;
    \item SMS management.
\end {itemize}

The UDM component works as a front-end for the user's subscription data recorded in the Unified Data Repository (UDR). UDM employs this subscription data to execute the logic of several applications, such as access authorization, registration management, and accessibility for finalizing events. UDR is a database where several types of data are stored, and the access is offered as a service to other components such as UDM, PCF, and NEF. There is an optional storage component called Unstructured data storage function (UDSF), which allows other components or functions to record dynamic context data outside the function or component itself. In the 3GPP context, unstructured data refers to the structure is not defined in the specifications, allowing each provider to use a give UDSF and choose its appropriate structure for storage. There is no requirement for any access compatibility or data storage in UDSF from different providers.

\subsection*{Network Repository Function (NRF)}

The NRF component is the repository, where all the functions available for a given network are listed. The goal of this component is to allow VNFs can find the appropriate function to meet their requirements. NRF has the responsibility to select the most suitable service provider component based on the performance criteria provided. In this way, the NRF component is updated whenever a new NBF instance is deployed or modified. Moreover, NRF holds information on the other VNFs, such as type, capacity, address, etc.

Based on the SBA scope, the NRF component plays a fundamental role in the functioning of the other VNFs. This component offers a central mechanism that allows automating the configuration process necessary for the other VNFs to discover and connect to specialized services.

\subsection{Additional Components}

In this subsection, we present additional components of the 5G core. These components are fundamental for creating a complete infrastructure of network functions for providing several business opportunities.

\subsection*{Policy Control Function (PCF)}

This component performs the same function as PCRF of the EPC core in the 4G system. PCF is responsible for controlling the behavior of the network, applying security and control rules, related to session management, mainly for functionalities associated with user mobility. This component interacts directly with AMF, providing an access and mobility policy that can add the control of access restrictions to services in a given area, for example. Moreover, PCF can include the management of topics associated with priority access to the channel of given UEs to the detriment of others. This management is named Radio Frequency Selection Priority (RFSP).

In the session management context, PCF can interact with the application functions and SMF. The main goal is to provide metrics associated with QoS and information regarding the data flow, which is obtained by regularly monitoring events related to the PDU session. Moreover, PCF offers security policy information for UEs. These policies can be associated with network resources and rules for selecting resource slicing. For example, PCF can be triggered to provide information when a given UE performs an access selection (UE access selection) or when a PDU session is established~\cite{hedman20195g}. The interaction between PCF and the other application functions is implemented through the exposure of six services, namely:

\begin{itemize}
\item \textit{PCF-AM-PolicyControl} -- provides information on access control policies, network selection, mobility management, and guidelines for the selection of routes between UEs and AMFs;
\item \textit{PCF-PolicyAuthorization} -- supplies authorization and provides access control policies to a request for an AF element, related to the PDU session in which AF is associated;
\item \textit{PCF-SM-PolicyControl} -- provides to SMF component guidelines for access related to PDU session in which the component is associated;
\item \textit{PCF-BDT-PolicyControl} -- supplies a set of guidelines for the NEF component to be used by applications to transfer data in the background plane;
\item \textit{PCF-UE-PolicyControl} -- provides control guidelines to be used in the communication process management between UEs and other network functions;
\item \textit{PCF-EventExposure} -- allows other network functions to register to be notified when a given event happens.
\end{itemize}

The decision on the application of the monitoring and control policies made by PCF is based, in part, on analytical information provided by other network functions, such as NWADF. Moreover, PCF is a fundamental component in a scenario where an AF needs to perform a given activity, \textit{i.e.}, data transfer in the background plane. In this case, AF can contact PCF to infer the best time interval for the activity's execution. This behavior allows the system operator to offer information to application providers about the most appropriate time to transfer data in the background plane.

\subsection*{Network Slice Selection Function (NSSF)}

In the context of the 5G networks, Network Slicing can be defined as a possibility for the system operator to allocate a resource set (in general, virtualized), to meet the requirements of given services or applications~\cite{7784887}. The aim is to offer support for a wide range of services, each with specific performance demands~\cite{foukas2017network}. The virtual resource slices are logical instances of network resources needed to meet a given request. A slice can include RAN and core resources, and even extrapolate to transport networks between several RANs and cores, considering network the end-to-end network service.

The 5G core architecture defines the NSSF component as being responsible for managing the available network slice instances. This component selects the network slice instances and the set of AMFs available for a given UE. AMF can be a component dedicated to a specific slice or to serve a set of network slice instances. The NSSF function is to assist AMF in choosing the available network slices, redirecting traffic between the controlled network slices. Moreover, NSSF can be seen as an orchestrator that can influence how network traffic is routed. This component produces two services, a selection service that provides information about the selected network slice and another availability service that generates information concerning the available resource slices.

\subsection*{Network Exposure Function (NEF)}

The NEF component is responsible for exposing some internal events related to UEs and the SBA model. The exhibition of these events aims to meet the demand for specific applications and VNFs of other components. For example, these demands need to access to the location or notify about the connectivity interruption of a given UE. Moreover, this information allows the AMF component to adjust the system according to user group behavior.

The possibility of exposing internal events through a NEF access interface opens new business opportunities for service providers, allowing in some cases, more advanced services to be offered by third parties. For example, an application can use the functions exposed by the NEF component to know if a given UE is accessible or not. Moreover, these functions can determine the geographical location of UE or if its device is in the movement. The NEF component responds to requests from several VNFs through regular interactions between UDM and AMF components.

\subsection*{Network Data Analytics Function (NWDAF)}

NWDAF is an optional component in the core architecture and is responsible for collecting several types of information from the network and its users. This information is later organized and analyzed to provide the inferred results for other network functions. The description of this component was written superficially in Release 15~\cite{3gpp:rel15nr21.915}, and its detailed specification is expected for Release 16.

The data collected by NWDAF comes from several other network functions that compose the core. The data collection is made through the services layer that connects the core components via the writing service. This service can be activated by the internal events triggered by each component. Moreover, NWDAF collects information about the operation of the system and data registry information in the UDR component.

Based on the specification, the services provided by NWDAF can be consumed by any core component. The external access is also possible using the NEF component. The analyzes made by NWDAF on the data collected over time can be used as a historical and statistical resource to predict future values. Moreover, the analytical data produced by NWDAF can be used to apply specific actions in the network context.

\subsection*{\textit{Application Function} (AF)}

AF is a generic component that represents a possible application, internal or external, to the operator's network, which interacts with the SBA model. The interaction process of AFs with SBA can influence some aspects of the whole system. For example, an AF can interact with the PCF component through services exposed by the NEF component, improving QoS aspects and, consequently, the charging and pricing policies.

An essential factor that must be evaluated by the system operator is the confidence degree that an AF component can have to interact directly with specific VNFs. For example, an AF with higher reliability can access VNFs from all components of the SBA model, directly. At the same time, a lower reliable AF must first interact with the NEF component before accessing more sensitive network functions. 

\subsection*{Non-3GPP InterWorking Function (N3IWF)}

The N3IWF component is used to integrate non-3GPP accesses with the 5G core. WiFi (IEEE 802.11) and Data Over Cable Service Interface Specification (DOCSIS) are examples of non-3GPP access technologies intended for integration by the standard. The conventional 3GPP access uses a BS, \textit{e.g.}, eNB (4G) or gNB (5G). However, the non-3GPP access starts on a different device, for example, a WiFi access point or a Hybrid fiber-coaxial (HFC) modem. This device uses the N3IWF component to access the 3GPP network and other 5G core components.

All traffic from the N3IWF component is sent through secure channels, and it is isolated from all 3GPP traffic. The isolation is maintained not only for data traffic (usual for 3GPP communications) but also for control traffic, including the traffic before the authentication process. More details about the N3IWF component, with its use and the interaction with other SBA components, are described in Subsection~\ref{subsec:nao-3GPP}.

\subsection{5G core demonstration}\label{sec:demo_nucleo}

\subsection*{Goals}

The main goal of this demonstration is to present in a practical way the main functionalities of the 5G core based on the SBA model. Moreover, we comment on the installation and configuration processes of software that implements the 5G core, providing material to replicate this demonstration.

\subsection*{Description}

\begin{figure*}[htb] 
 \begin{center}
\includegraphics[width=0.6\textwidth]{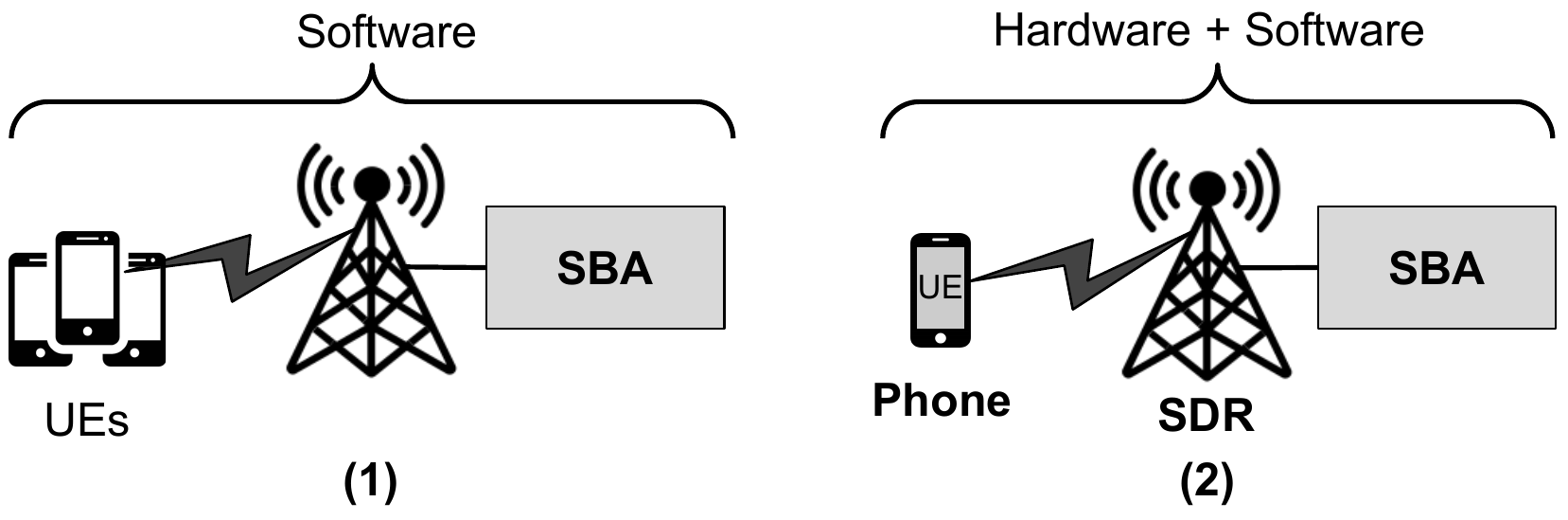}
  \end{center}
\caption{Experiments with a 5G core.}
\label{fig:demo2}
\end{figure*}

In this demonstration, we use the 5G core based on the open-source SBA model. We also present how to emulate RAN and multiple UEs. The presentation is organized in two experiments: (1) UEs, RAN, and core, all implemented in software; (2) UE in hardware (conventional mobile phone), eNodeB in hardware (SDR) and software, and the 5G core implemented in software. All components are implemented using Docker containers that can be hosted on a cloud infrastructure. Fig.~\ref{fig:demo2} shows the software and hardware components that are used in the 5G core demonstration.

\subsection*{Additional information}

During the tutorial, we present demonstration videos of the experiments. Moreover, the manuals are available with details on how the practices can be replicated. Finally, containers and any extra source code produced by the team and need to replicate the experiments are publicly available.

Repository for this tutorial:\\
\url{https://github.com/LABORA-INF-UFG/NetSoft2020-Tutorial4}.

\section{Integration of 5G system with non-3GPP access networks}\label{sec:integracao}

Although 5G networks have been designed to serve a wide variety of services, offering a board range of resources, 3GPP envisaged in Release 15 the integration and support of other non-3GPP wireless access technologies, with particular attention to WiFi (IEEE 802.11). As already discussed in some investigations \cite{parkvall2017nr}, \cite{ghosh20195g}, the coexistence of different access technologies promotes significant gains concerning the performance and cost of the communication of IoT devices. Moreover, many devices have a non-3GPP wireless interface, and these devices will be in operation for some years. The integration of multiple wireless access technologies in an only core can benefit both users and administrators of public and private communication infrastructure.

Release 15~\cite{3GPP:19} already defines how non-3GPP wireless network technology can be used as access and integrated into a 5G core. However, this release focus only on Wireless Local Area Networks (WLANs), \textit{i.e.}, WiFi networks, considering the untrusted access. The trusted access will be introduced in Release 16~\cite{WBA-NGMN:19}. In this release, attention is also being taken to cable access technologies, in particular for HFC. Although 3GPP standards do not explicitly describe how to integrate other wireless communication technologies used in IoT, such as LoRa and ZigBee, the framework allows extensions to be exploited~\cite{navarro-ortiz:18,yasmin:17}. In the following, we present more details about Release 15 and also on LoRa/LoRaWAN technology, which we use as an illustration for the integration between the 5G core and a non-3GPP wireless access technology at the end of this section.

\subsection{Non-3GPP Access Networks}\label{subsec:nao-3GPP}
 
The 5G core architecture foresees, since its design, the possibility of integrating non-3GPP access technologies, with the premise that the communication interfaces offer IP connectivity, \textit{i.e.}, a traditional IP stack. Although this premise is met for technologies such as WiFi and HFC, several IoT solutions such as LoRa, ZigBee, nRF24, and others consider that exists a gateway to provide the IP connectivity for the IoT devices. Therefore, we assume from this point in the article that equipment with a traditional IP stack, \textit{i.e.}, a device or a gateway, is available to establish the integration with the non-3GPP network. It is essential to highlight, as describe by~\cite{yasmin:17} that there are several ways to integrate devices that depend on a gateway with a traditional IP stack.

The EPC core from 4G also has support for the non-3GPP access network, using a more sophisticated approach. Moreover, for defining between trusted and untrusted access, it needs to choose between network-based mobility and host-based mobility. Considering the focus of this article, no further information on the EPC core approach will be present, but a full description can be found in~\cite{olsson:13}. For the 5G SBA core, fewer options were defined, basically, \textbf{untrusted} non-3GPP access (Release 15, frozen in March 2019 and completed in June 2019) and \textbf{trust} (Release 16, expected to freeze in March 2020 and conclusion to June 2020). In this context, the term untrusted means that the 3GPP network operator does not trust in the security offered by the non-3GPP access network. Therefore, it needs to take actions that ensure the proper transport of traffic from this access network. This need means that the non-3GPP traffic must be isolated from other traffics, including the 5G core, which is suitable for IoT applications and services. The trust access does not allow integration with other wireless communication technologies used in IoT, such as LoRa or ZigBee.

\begin{figure}[htb]
 \begin{center}
\includegraphics[width=0.5\textwidth]{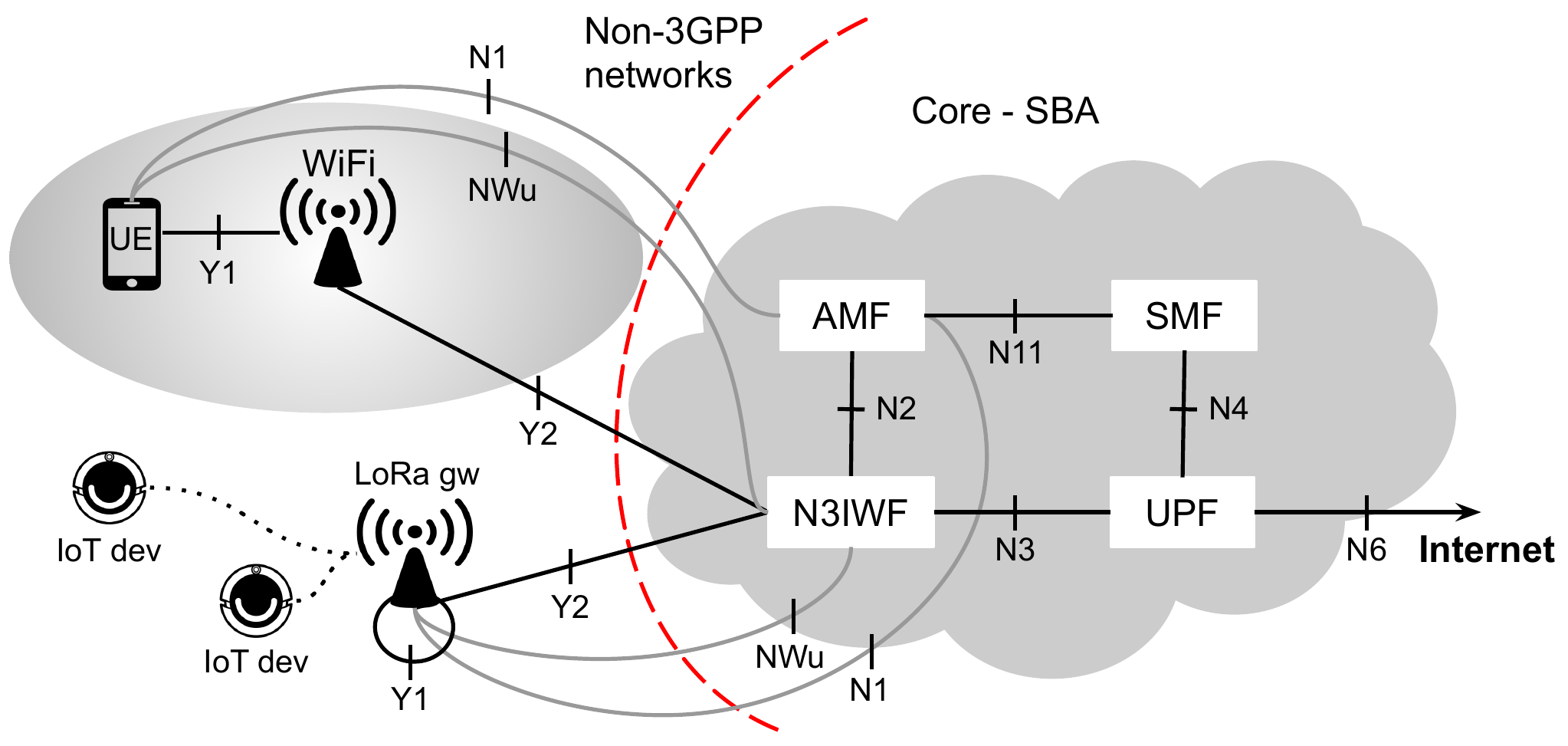}
  \end{center}
\caption{Integration between the untrusted non-3GGP access networks and the 5G SBA core.}
\label{fig:non3gpp_untrusted}
\end{figure}

To support the untrusted non-3GPP access network, the main component introduced by 3GPP Release 15 was N3IWF, which is responsible for forwarding signaling and data between the 5G core and the non-3GPP access network, as described in Subsection \ref{sec:core}. Fig.~\ref{fig:non3gpp_untrusted} illustrates the integration of two non-3GPP access networks to a 5G core, showing the main components involved and their communication interfaces. N3IWF selects AMF to serve the IoT device (or gateway), which will be responsible for holding the (eventual) mobility of the equipment and brokering all the signaling with other 5G core functions. For effective communication, a device needs PDU sessions, which are established, modified, and released under the SMF component's control. This component works as a CP that operates on a DP implemented through the UPF component.

\begin{figure}[htb]
 \begin{center}
\includegraphics[width=0.5\textwidth]{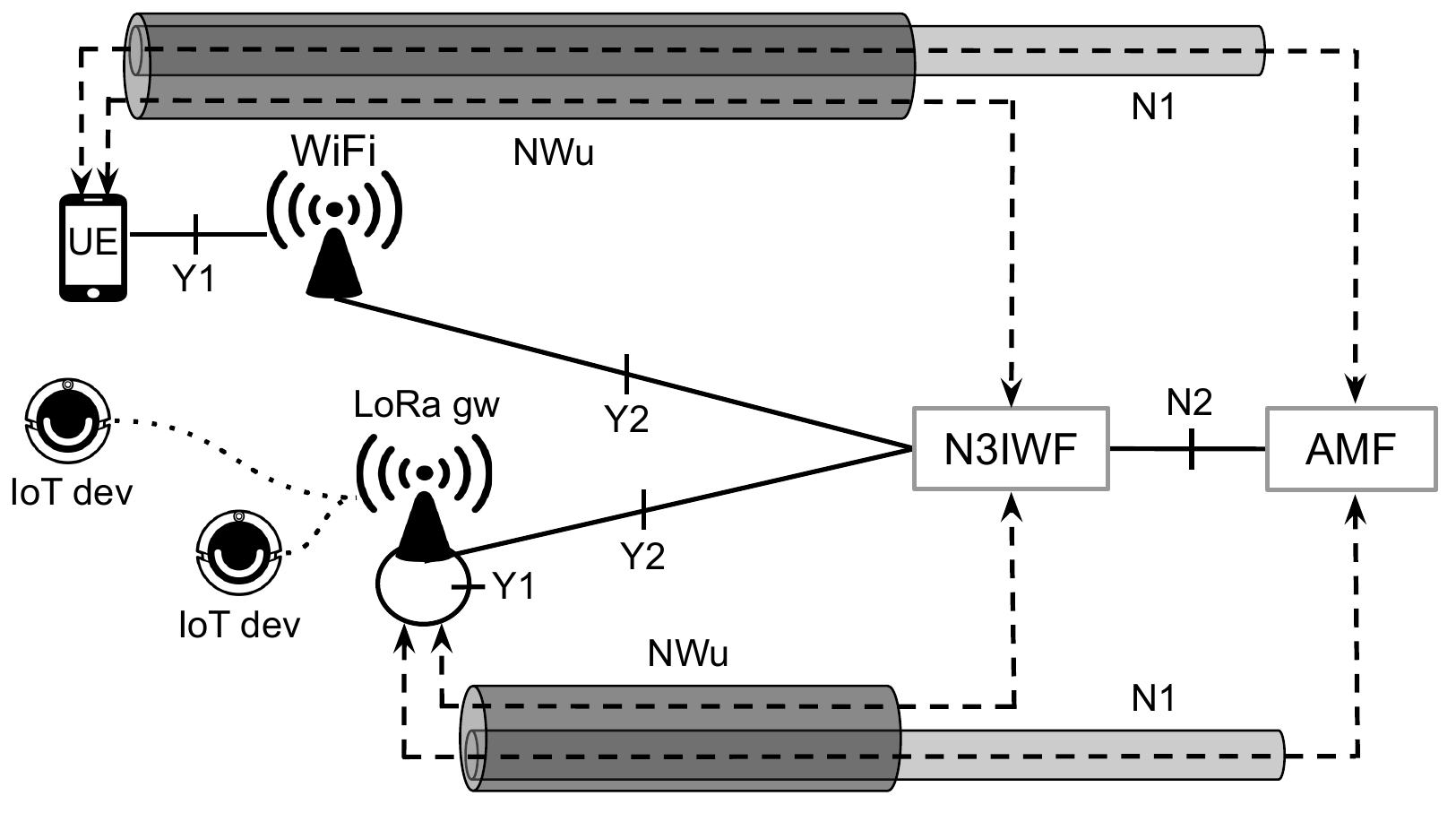}
  \end{center}
\caption{Connections for the integration of untrusted non-3GPP access network with the 5G SBA core.}  
\label{fig:non3gpp_connections}
\end{figure}

In the 5G SBA core, the AUSF component allows a device to authenticate itself to access the network and its services using a 3GPP wireless interface (\textit{e.g.}, NR) and a non-3GPP interface (\textit{e.g.}, WiFi). Non-3GPP devices can authenticate with the SBA core through a certification-based scheme, using Extensible Authentication Protocol-Transport Layer Security (EAP-TLS) or EAP-Tunneled TLS (EAP-TTLS) and the traditional procedure with credentials based on Subscriber Identification Module (SIM). Fig.~\ref{fig:non3gpp_connections} shows the connections established to provide integration between the untrusted non-3GPP access network and the 5G SBA core.

3GGP does not define the Y1 connection, but this connection establishes the communication of the device with its access network, \textit{e.g.}, through a WiFi access point or LoRa gateway. It is assumed that there is some authorization process through the Y1 connection, allowing the device to access the network and to obtain an IP address. However, several technologies, such as LoRa, do not directly assign IP to the devices, and part of the Y1 connection needs to be implemented at the technology gateway itself. The Y2 connection is also not defined by 3GPP, but it must provide communication between the access network and the N3IWF component of the SBA core. This connection can be established in almost any way, including via the public Internet, \textit{i.e.}, potentially with multiple equipment and intermediate technologies.

The 3GPP standard specifies that between the device (or gateway) and the N3IWF component, an encrypted IPSec tunnel, called NWu, must be established for sending data and signaling traffic. N3IWF selects AMF to serve the device, and the N2 interface between the two functions is established. After setting NWu and N2, the 3GPP standard specifies that it is possible to create the N1 interface between devices (or gateway) and the AMF component for transporting Non-Access Stratum (NAS) signaling. This approach differs from that used in the 4G EPC core, in which NAS signaling applies only to 3GPP access networks. In the 5G SBA core, devices connected via non-3GPP access networks are managed similarly to devices connecting via 3GPP access networks.

\subsection{LoRa}

LoRa technology, developed by Semtech Corporation~\cite{semtech}, allows communication over long distances using the Spread Spectrum concept of RF. This technology has the following configuration parameters that directly influence the communication performance:

\begin{itemize}
\item Spreading Factor (SF) with values of 7, 8, 9, or 10. The higher SF, more information is transmitted per symbol, generating a gain in transmission;
\item Bandwidth (BW) of 125 kHz, 250 kHz, or 500 kHz for a given SF. A narrower BW increases the reception sensibility and increases the packet transmission time;
\item Code Rate (CR) is responsible for detecting and correcting errors.
\end{itemize}

These configurations determine the bit rate transmitted, the maximum size of the transmitted data, and the transmission time of a packet in the RF spectrum. These configurations also influence the size of messages, their range, and the energy consumption of the IoT device.

According to ~\cite{shanmuga2020} and ~\cite{Lee2017}, Lora is resistant to interference, due to its wide range of SFs, which can be used in urban, rural, and even industrial environments. Moreover, this technology offers four times the coverage range, compared to other radio technologies due to its robust nature and ability to work with low-intensity radio signals. Its ability to perform multiple transmissions on the same RF channel, with the use of different SFs, reduces the likelihood of collisions, increasing the effective data transmission rate, allowing discrimination between time and frequency errors. Therefore, LoRa is considered one of the most promising technologies for the physical layer of sensor networks and, consequently, for IoT~\cite{Haxhibeqiri2018}. However, to have a useful application of sensors and IoT, it is necessary to have a network, such as LoRaWAN, described below.

\subsection{LoRaWAN network}

\begin{figure}[!h]
 \begin{center}
\includegraphics[width=0.45\textwidth]{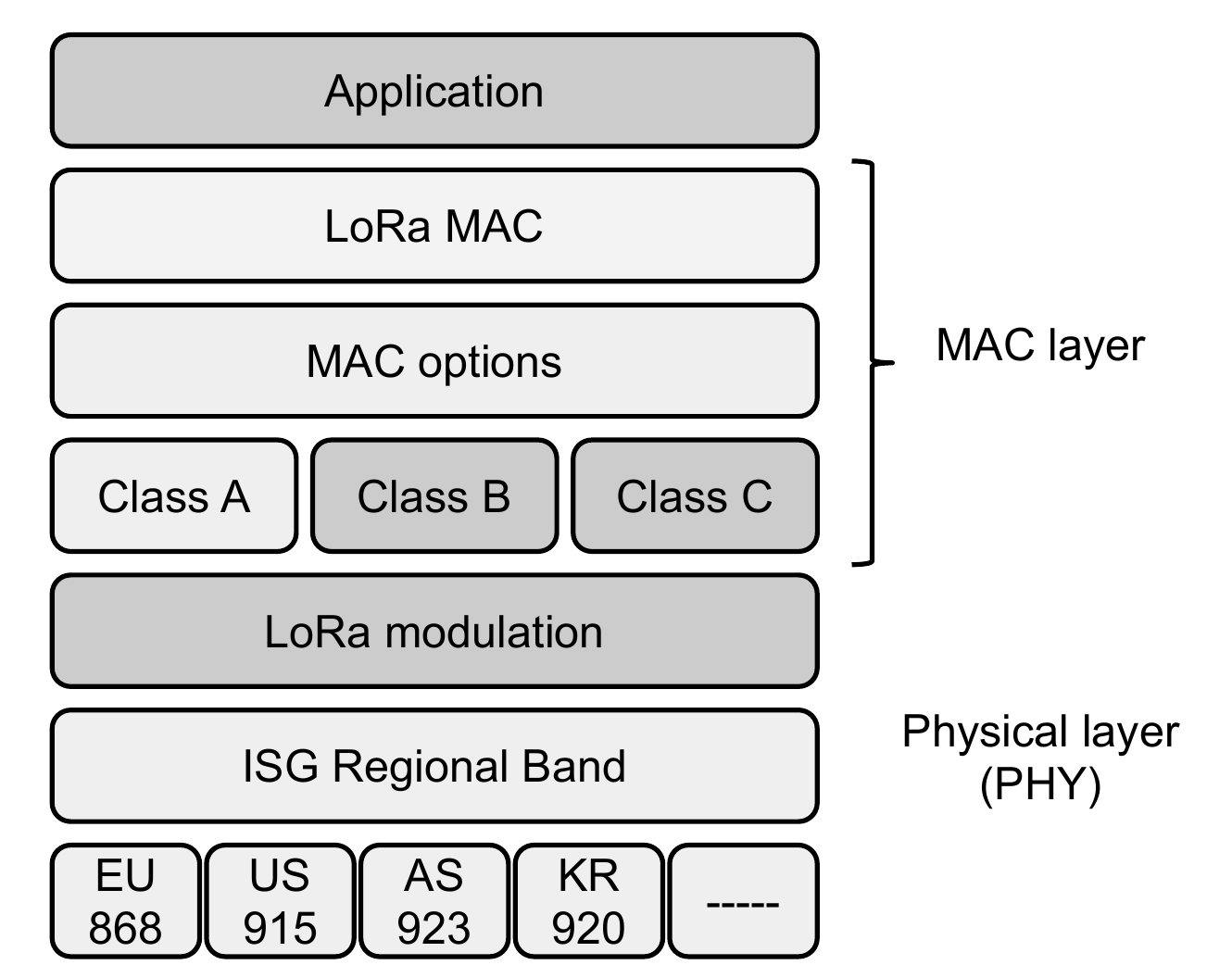}
  \end{center}
\caption{LoRaWAN architecture layers.}
\label{fig:mac_lorawan}
\end{figure}

LoRaWAN network is the name given to IoT Low Power Wide Area Network (LPWAN), which uses the LoRa technology as a physical medium and implement an architecture based on the LoRaWAN Media Access Control (MAC) protocol. In Fig.~\ref{fig:mac_lorawan}, we can see the MAC layer, its sublayers, and the LoRa physical layer, enabling the long-range communication link.  The MAC layer protocol and the network architecture have a significant influence on determining the battery life a sensor, network capacity, QoS, security, and the range of IoT applications served.

The LoRaWAN protocol is based on open, low-cost standards. This protocol was designed from the beginning to implement platforms for IoT. In this way, the scope of supported IoT applications is broad, making this technology attractive to the 5G system and creating an IoT network with a large coverage area. Among the possible uses, there are smart grids for electrical energy, sensor networks of several types, precision agriculture, and others.

\begin{figure}[!h]
 \begin{center}
\includegraphics[width=0.5\textwidth]{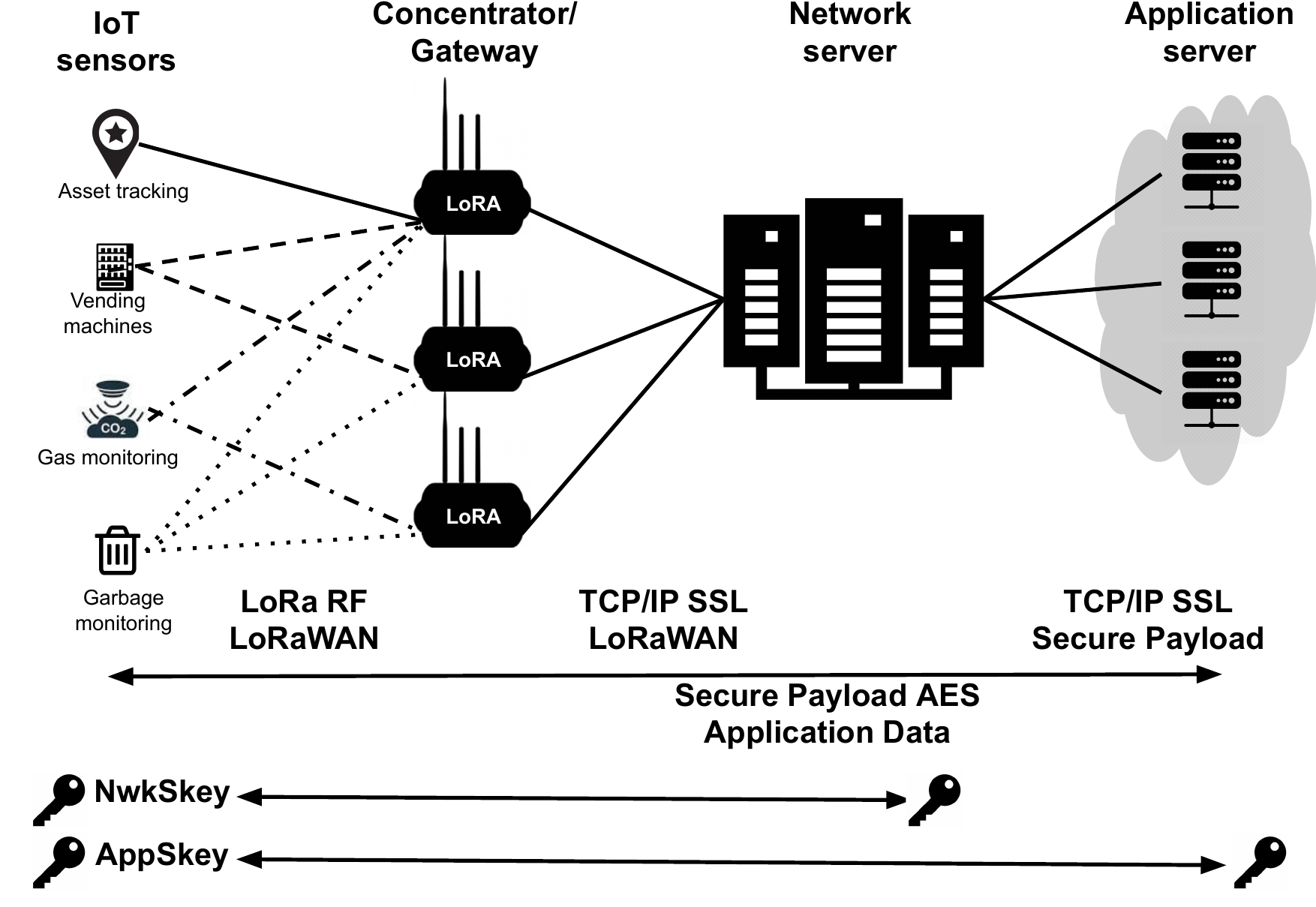}
  \end{center}
\caption{LoRaWAN network topology}  
\label{fig:topologia_lorawan}
\end{figure}

Fig.~\ref{fig:topologia_lorawan} shows the LoRWAN protocol reference topology. A prominent feature of this technology is its low energy consumption, due to the star topology that does not require routing between nodes. The central component is the LoRaWAN gateway that can be connected to a 5G system on untrusted non-3GPP access. These gateways, also known as concentrators, have models that can meet specific demands, simpler and cheaper for closed environments, in contrast to industrial models with protection against the weather and external applications. In communication between gateway and devices, there are two previous types of messages exchange: Unconfirmed Data Message, similar User Datagram Protocol (UDP) messages, and Confirmed Data Message, like to Transmission Control Protocol (TCP). Another fundamental characteristic of the LoRaWAN network is the existence of three devices classes that communicate with the gateways, which are described in the following.

\begin{itemize}
\item Class A: upstream transmissions (device-gateway) are based on the ALOHA protocol. The reception of the can only be performed in two short reception windows that open after a transmission. This class provides the lowest energy consumption, with the main application of the monitoring quantities.
\item Class B: in addition to the class A mode, pre-programmed transmission windows (gateway-device) and managed by a timing signal (Beacon) are opened, which indicates when the receiver is ready to receive the signal. This class can be convenient for remote control systems that are not time-sensitive.
\item Class C: the transmission window (gateway-device) is open continuously, closing only at the time of the upstream transmission (device-gateway).
\end{itemize}

Regarding the activation of devices on the network, LoRaWAN adopts the AES-128 encryption algorithm, using two different forms of activation due to the Public or Private network. In public networks, it uses Over The Air Activation (OTAA), which is based on sending a unique global identifier (DevEUI), analogous to the MAC address, identifier (AppEUI) and application key (AppKey) desired. This data is used in the application layer to validate and activate it in a given network application. After the acceptance on the network, the device receives a message \texttt{join accept}, which has the device address (DevAddr), the network session key (NwkSKey), and the application session key (AppSKey). In private networks, the following data are required for activation: device address (DevAddr), network session key (NwkSKey), and application session key (AppSKey), which are recorded on the device, at the time of configuration. Therefore, the device is ready for communication when it is connected to the network.

In the LoRaWAN architecture, Network Servers are still responsible for managing the information sent by gateways. As there is the possibility that two or more gateways receive the same packet from a given device, the network server eliminates duplicate packets, manages the acknowledgment (ACK) return times, and makes the adjustments for the adaptative data rate to manage the times between communications and energy consumption. Finally, LoRaWAN has one or more Application Servers that receive packets from Network Servers via request or automatically and, according to the request, perform one or more specific actions providing the needed interface for various client applications. 

\subsection{Demonstration of the integration of non-3GPP wireless access network and the 5G core} \label{sec:eRAN}

\subsection*{Goals}

The goal of the demonstration is to introduce a non-3GPP wireless access network to the 5G core. Moreover, this experiment shows the great capillarity that the 5G system can have, including network with unlicensed frequencies. 

\subsection*{Description}

\begin{figure}[htb] 
 \begin{center}
\includegraphics[width=0.5\textwidth]{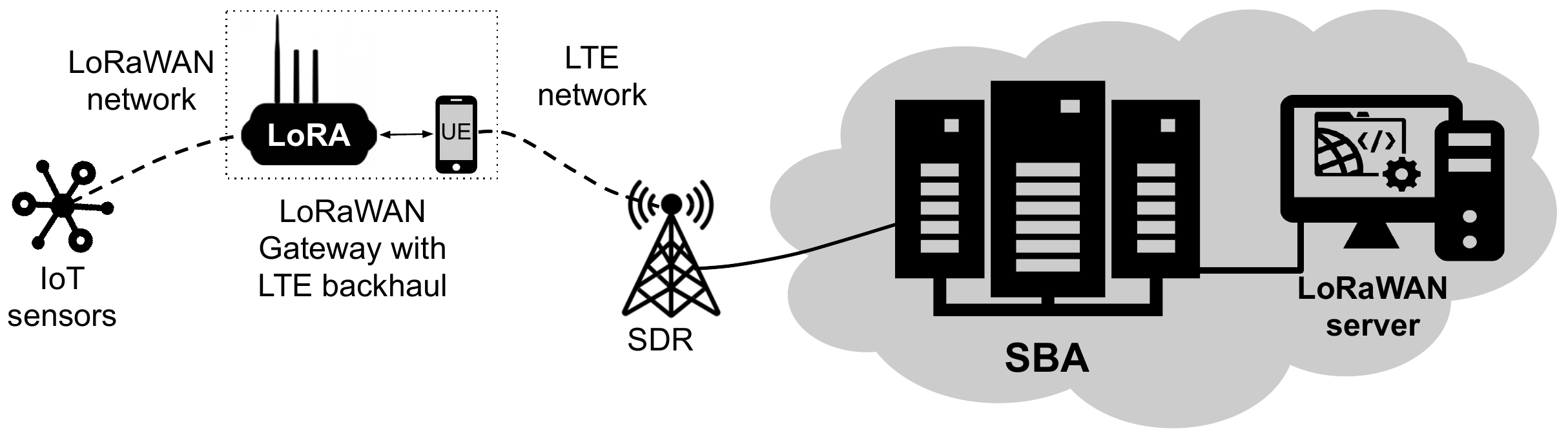}
  \end{center}
\caption{Demonstration of the integration of a 5G core with a non-3GPP access network based on LoRa.}
\label{fig:demo3}
\end{figure}

In this demonstration, we combine a RAN based on LTE technology with a LoRaWAN wireless network implemented in hardware. For RAN LTE, we use open-source software and an SDR. We also use an open-source implementation of the SBA-based 5G core software. All components are implemented using Docker containers that can be hosted on a cloud infrastructure. Connected to access the network through LTE, there is a LoRaWAN gateway implemented in generic hardware with support to multiples IoT sensors that synchronize its data with the LoRaWAN server present at the other end of the infrastructure. Therefore, for demonstration purposes, the data collected by the sensors are transmitted via the LoRa network to the gateway that forwards the data via LTE backhaul, passing through the SBA core of the 5G network until reaching the LoRaWAN server. Fig.~\ref{fig:demo3} shows the experiment we designed to demonstrate the integration of a non-3GPP wireless access network with a 5G core network.

\subsection*{Additional information}

During the tutorial, we present a demonstration video of the experiment. Moreover, a manual is available with details on how the practices can be replicated. Finally, containers and any extra code produced by the team needed to replicate the experiment is also publicly available.

Repository for this tutorial:\\
\url{https://github.com/LABORA-INF-UFG/NetSoft2020-Tutorial4}.

\section{Final considerations}\label{sec:conclusao}

The academy has already started to investigate potential challenges for the next generation of wireless mobile networks. Naturally, there is still a lot of uncertainty, and several forecasts cannot be consolidated. However, some themes are beginning to be addressed preliminarily in the 3GPP release scope. For example, Releases 16 and 17 improvements in URLLC and the use of SON to make the 5G networks more efficient are expected, indicated that forecasts such as MBRLLC and mURLLC, as well as, the broad adoption of Machine Learning and Artificial Intelligence, can become a reality in the future. In the following, we briefly discuss some topics addressed in the next two releases, scheduled for June 2020 (Release 16) and September 2021 (Release 17).

\subsection*{Data-driven network}

The introduction of NWDAF in Release 15 is an essential step for adopting ML and AL in a 5G network, but it is only the beginning. The complete specification of the data collection and analysis framework must be completed in Release 16. From that point, solutions based on ML and AI can use the information collected by NWDAF to perform tasks such as predicting and mobility optimization, detecting anomalies, QoS predicting, and data correction. In this context, some of the goals for Release 17 are:

\begin{itemize}
    \item Predictable network performance assisted by NWDAF;
    \item UE oriented data analysis;
    \item Expose of NWDAF data analysis for user applications;
    \item NWDAF supporting the detection of anomalous events and helping to analyze their causes.
\end{itemize}

In the long term, the objective is to use ML and AI techniques to automate network management with the minimum possible human intervention. It is fundamental to highlight that this management should involve different network types (\textit{i.e.}, heterogeneous networks) connecting to a common core based on the SBA model.

\subsection*{Improvements for vertical domains}

5G networks have been designed with particular attention to vertical domains (usually called only as verticals), \textit{i.e.}, specific sectors or groups of companies in which similar products and services are developed, produced, and supplied. In this context, there are several directions and contributions in Release 16, for example:

\begin{itemize}
    \item Support to Time Sensitive Communication (TSC);
    \item Non-Public Networks (NPNs), \textit{i.e.}, private networks;
    \item Support to 5G LAN services;
    \item Advanced location services.
\end{itemize}

In Release 17, improvements in support for NPNs and Proximity Services (ProSe) are planned, which are also useful in public security scenarios.

\subsection*{Security evolution}

Security has always been a concern in mobile wireless networks, especially from 3G with the wide use of the infrastructure for data transport and its natural integration with the Internet. Therefore, the security issue in 5G networks has been addressed from its design. It has been a recurring topic, mainly due to the strategic importance of 5G in the economy and social issues. There is a list of security contributions planned for Release 16, of which some relevant items are highlighted the following:

\begin{itemize}
    \item Support for NPNs with new authentication schemes;
    \item Network slice specific authentication option, in addition to primary authentication;
    \item Advanced security for Radio Resource Control and NAS signaling;
    \item Support for integrity protection in the user plane, covering the three 5G network scenarios, \textit{i.e.}, eMBB, URLLC, and mMTC.
\end{itemize}

\subsection*{Other advances}

There are several other contributions in Release 16 and 17 that introduce new solutions and offer opportunities for the research. This opportunity occurs because specific solutions are not entirely defined due to factors such as complexity of the problem addressed or inability to identify in the standard the specifics that the answer will have to deal with that problem. In the following, we list some topics of these two Releases (16 and 17) that, although they have already been investigated, still present open questions, especially in a practical context and with the potential large-scale application. 

\begin{itemize}
    \item Advances on Vehicle-to-everything (V2X) -- platoon formation, autonomous steering, and remote steering;
    \item Access to unlicensed spectrum using 5G New Radio; 
    \item Dynamic Spectrum Sharing (DSS);
    \item IoT for Non-Terrestrial Networks (NTNs);
    \item Support for Unmanned Aerial Systems (UAS).
\end{itemize}

Finally, we can see the increasingly strong integration between the areas of Telecommunications and Information Technology. On the one hand, the adoption of software and cloud-native model in 5G systems, \textit{i.e.}, RAN and core has grown. On the other hand, the interest to create software solutions that meet the specific needs of mobile wireless networks, such as edge computing and signal processing systems, has also increased. This integration will still be explored mainly in the next years, to make development even more agile and interoperable. An example of this change in view is the distribution (from Release 15) of 3GPP specifications in the format of readable data serialization (YAML - \textit{YAML Ain't Markup Language})~\cite{3GPP:OpenAPI}, allowing the design and development of standardized interfaces and with high flexibility. Moreover, the use of cloud-native technologies will allow extremely automated management and control, with minimal (or even without) human intervention, which has been called Zero Touch Network $\&$ Service Management (ZSM)~\cite{etsi:ZSM}. These developments are expected to change the business model of telecommunication operators, allowing them to expand Business to Consumer (B2C) and Business to Business (B2B) relationships.

\bibliographystyle{IEEEtran}


\end{document}